\documentstyle[amssymb,aps]{revtex}

\begin{document}
\author{Edward Gerjuoy}
\address{Dept. of Physics, University of Pittsburgh,\\
Pittsburgh, PA 15260}
\title{Shor's Factoring Algorithm and Modern Cryptography. An Illustration of the
Capabilities Inherent in Quantum Computers }
\date{5/1/04}
\maketitle

\begin{abstract}
The security of messages encoded via the widely used RSA public key
encryption system rests on the enormous computational effort required to
find the prime factors of a large number {\it N }using classical (i.e.,
conventional) computers. In 1994, however, Peter Shor showed that for
sufficiently large $N$ a quantum computer would be expected to perform the
factoring with much less computational effort. This paper endeavors to
explain, in a fashion comprehensible to the non-expert readers of this
journal: (i) the RSA encryption protocol; (ii) the various quantum computer
manipulations constituting the Shor algorithm; (iii) how the Shor algorithm
performs the factoring; and (iv) the precise sense in which a quantum
computer employing Shor's algorithm can be said to accomplish the factoring
of very large numbers with less computational effort than a classical
computer can. It is made apparent that factoring $N$ generally requires many
successive runs of the algorithm. The careful analysis herein reveals,
however, that the probability of achieving a successful factorization on a
single run is about twice as large as commonly quoted in the literature.  
\end{abstract}

\section{INTRODUCTION.}

Recently published papers in this journal$^{1,2}$ have attempted to explain
how a quantum computer differs from a classical (i.e., a conventional)
computer. However neither of these papers offers a detailed discussion of
the factoring algorithm developed by Peter Shor$^{3}$ in 1994 although this
algorithm, whose implementation could frustrate one of the most widely used
modern methods of encrypting messages, provides the most impressive known
illustration of the increased computing power potentially attainable with
quantum computers. The present paper seeks to fill the gap, namely it seeks
to furnish, self-contained in this journal's pages but with full citations
to the relevant literature, a comprehensible explanation of Shor's
algorithm, as well as of the algorithm's relevance to modern cryptography.
In so stating I am aware that although Shor's original paper$^{3}$ and other
available publications providing full discussions of the Shor algorithm
typically are written for quantum computing specialists$^{4}$, less
technical presentations$^{5}$ more suitable for the non-specialist readers
of this journal can be found. I additionally remark that various Internet
websites$^{6}$ post links to a wide variety of publications in the quantum
computing literature, organized under numerous suitable headings including
Shor's Algorithm.

I now summarize the contents of this paper. The immediately following
Section II first describes the basic elements of classical cryptography,
wherein {\it keys} are employed to encipher and/or decipher messages in
order to prevent those messages from being read by anyone other than their
intended audience. Section II then explains in detail, and also explicitly
illustrates, the enciphering and deciphering procedures in the so-called 
{\it RSA system}$^{7}$, an important modern scheme for sending secret
messages. These explanations and illustrations necessarily involve
presentations of the number theory underlying the RSA system; without some
understanding of this underlying number theory the ability of the RSA system
to transmit secure messages seems magical. Section II goes on to explain how
critically the security of the RSA system depends on the fact that using
classical computers to factor large numbers requires huge outlays of
computer resources and time; the relevance of Shor's factoring algorithm to
the security of the RSA system thereby becomes evident. The next Section
III, after briefly summarizing the relevant properties of quantum computers,
fully describes and illustrates Shor's algorithm. Section III also explains
the precise sense in which a quantum computer employing Shor's algorithm can
be said to accomplish the factoring of very large numbers with less
computational effort than a classical computer can. To avoid unduly
interrupting the flow of the discussion, many details of the underlying
number theory, which are important not only for understanding why the RSA
system works but also for comprehending how Shor's algorithm enables the
factorization of large numbers, are relegated to the final Section IV which
serves as an Appendix.

Before closing this Introduction I emphasize that except for its discussion
of the Shor algorithm, which is designed for a quantum computer, this paper
is concerned solely with classical computers. In particular I assume that
the RSA system enciphering and deciphering described in Section II are
performed with classical computers, as experience has demonstrated is
practical with numbers of the size presently being used as RSA keys. The
possible use of quantum computers to perform such enciphering and
deciphering, and any other aspects of quantum cryptography and/or
computation aside from Shor's algorithm, are beyond the scope of this paper.
Also beyond the scope of this paper are the difficulties, observed and
anticipated, involved in actually constructing functioning quantum
computers; such difficulties are amply discussed in the literature, e.g., in
various chapters of Nielsen and Chuang.$^{4}$

\section{Cryptography, Key Distribution and Number Theory.}

As Ekert$^{8}$ observes, ''Human desire to communicate secretly is at least
as old as writing itself and goes back to the beginnings of our
civilization.'' The full history of secret communication until about 1965 is
recounted by Kahn$^{9};$ developments after about 1965, including those
advances in secret communication to which Shor's algorithm pertains, are
described by Singh$^{10},$ who also (but less fully than Kahn) reviews the
pre-1965 history. Kahn$^{9}$ carefully defines the terms {\it plaintext}
(the original uncoded message), {\it cryptogram} (a writing in code, e.g.,
the enciphered message), {\it key }(the information or system employed to
encipher the plaintext message), {\it cryptography} (the acts of enciphering
the plaintext into a cryptogram, and/or of deciphering the cryptogram by
someone who knows the key), and {\it cryptanalysis} (the art/science of code
breaking, i.e., of ferreting out the key from the cryptogram); this paper
adopts Kahn's terminology.

Until about 1975 cryptographers employed so-called {\it symmetric} or {\it %
private }(also known as {\it secret)} key systems only.$^{11}$ In such
systems the key used by Alice (conventionally the sender of the cryptogram)
to encipher the message is the same as the key Bob (conventionally its
receiver) employs to decipher the cryptogram. One of the simplest type of
symmetric keys (which also appears to have been the earliest type to be
employed, dating back to nearly 2000 B.C.$^{12}$) is termed {\it substitution%
}. In substitution-key cryptography the cryptogram is constructed from the
plaintext by replacing each letter (of the alphabet) in the plaintext with
some other chosen expression; the replacement expression can be another
letter of the alphabet, or a symbol of some kind, or an arbitrary
combination of letters and symbols; the same letter of the alphabet in
different portions of the plaintext may be replaced by different
expressions; the key is the chosen replacement scheme. In the very simplest
substitution keys, which for the purpose of this paper may be termed {\it %
unique substitution}, Alice and Bob agree that the replacements will be
unique and one to one, i.e., (i) that any given letter in the plaintext
always is replaced by the same expression, and (ii) that different letters
of the alphabet are replaced by different expressions. Though cryptograms
constructed via unique substitution keys can seem impregnable, especially
when the key involves unusual or unfamiliar symbols, they actually are
readily deciphered taking advantage of the peculiarities of the language
(assumedly known or correctly guessed) in which the plaintext had been
written, as 15th century Arab cryptologists already knew.$^{13}$ The
writings of Edgar Allan Poe$^{14}$ and Arthur Conan Doyle$^{15}$ provide
celebrated fictional cryptanalyses of unique substitution cryptograms (in
these cases with English plaintexts) wherein letters of the alphabet had
been replaced by symbols. In the most transparent unique substitution
cryptograms a single alphabet letter is{\it \ }substituted for each
plaintext letter, consistent with a preselected so-called {\it cipher
alphabet;} cryptograms constructed in this fashion, but employing a
different cipher alphabet each day, are regularly published in many daily
newspapers$^{16}$ as puzzles to be deciphered by the newspaper's readers
using, e.g., the fact that in English the letter $e$ is by far the most
frequent.$^{17}$

But substitution cryptograms need not be constructed with unique keys;
moreover nonunique substitution cryptograms can be and have been made very
difficult to cryptanalyse. A famous illustration of this last assertion is
provided by the Enigma machine employed by the German army during World War
II, which constructed cryptograms wherein each letter was replaced by a
single letter as in newspaper cryptograms, but wherein the cipher alphabet
employed to encipher any given plaintext varied not merely from day to day,
but also from one plaintext letter to the next, in accordance with a
predetermined randomly selected complicated key.$^{18}$ Whether or not very
difficult to cryptanalyse, however, all substitution and other symmetric key
cryptographic systems have a deficiency known as the {\it key distribution
problem, }as numerous authors have observed$^{19}$: Before Alice and Bob can
begin exchanging hopefully non-cryptanalysable cryptograms, they must
exchange--in a non-encrypted form--the information necessary to establish
their key; they cannot be confident this information exchange has not been
intercepted unless the exchange takes place in the same room, and perhaps
not even then.$^{20}$ Symmetric key cryptographic systems also have the
related deficiency that if Bob wants to receive secret messages from more
than one Alice, then he either must set up different keys with each Alice or
else risk the possibility that one Alice will intercept and readily decipher
a message sent Bob by another Alice.

These deficiencies are avoided in {\it asymmetric} (commonly termed {\it %
public}) key systems, wherein the key Alice employs to encipher her message
is not the same as the key Bob employs to decipher the cryptogram he
receives. The differences between symmetric and asymmetric key systems can
be visualized in terms of a safe: With a symmetric system the key Alice
employs to open the safe and lock her message inside it is the same as the
key Bob employs to open the safe and remove the message. With an asymmetric
system Alice's key enables her to open the safe just enough to insert her
message, but no more; only Bob's key can open the safe's door sufficiently
widely to permit message removal.$^{21}$ Indeed in the RSA system$^{7},$ one
of the most commonly employed public key systems, Alice's enciphering key is
made public, i.e., is not at all secret but rather is available to any Alice
who wishes to send Bob an encrypted message; Bob's deciphering key remains
his secret. The RSA system is detailed immediately below.

\subsection{The RSA Public Key System.}

.Bob creates his RSA public key system in the following fashion$^{7}$: He
first selects two different large prime numbers {\it p} and {\it q,} and
then computes their product {\it N = pq. }Next he also computes the product 
{\it \ }$\phi $ {\it = (p-1)(q-1) = N +1-(p+q), }and then selects a positive
integer $e$ which is coprime to{\it \ }$\phi $ (meaning {\it e} and $\phi $%
{\it \ }have no common prime factors other than 1). Finally he computes a
positive integer {\it d} such that the product {\it L = de }has the
remainder unity when divided by $\phi .${\it \ }Bob now has all the required
components of his public key system, except that because the procedure by
which messages are encrypted in his system{\it \ }involves arithmetical
manipulations of integers (as I am about to explain), the system also must
include some specified{\it \ }convenient-to-use symmetric key whereby any
Alice can convert her plaintext into a cryptogram consisting of a sequence 
{\it C} of positive integers {\it c,} which sequence she then will further
encrypt (via Bob's proclaimed procedure) into the sequence {\it S }of
positive integers $s$ actually sent Bob. {\it \ }

The non-secret components of Bob's public key system, which Bob now is ready
to broadcast for the benefit of one and all, are$^{7}$: (i) the positive
integers {\it N} and{\it \ e}; which herein will be termed the {\it key
number} and {\it encryption exponent} respectively; (ii) the details of the
symmetric key Alice will be using to construct her $C,$ and which Bob also
will use to reconstruct Alice's original plaintext message once he has
deciphered $S$ and thereby recovered $C$ (the only restriction on {\it C }is
that every {\it c} must be less than $N)$; and (iii) the surprisingly simple
procedure for constructing the elements {\it s} of $S$ from the elements $c$
of $C$, namely $s$ is the integer remainder when {\it c}$^{e}$ is divided by 
$N.$ Note that the symmetric key which Alice and Bob share now is completely
public; there is no attempt whatsoever to keep it secret. Similarly Alice
transmits her finally enciphered cryptogram $S$ to Bob via perfectly open
communication channels, e.g., by email. The secret procedure by which Bob
extracts the original $C$ from $S$ parallels, but is not the direct inverse
of, the public procedure which constructed $S$ from $C,$ namely$^{7}$ for
each $s$ Bob computes the integer $u$ which is the remainder when $s^{d}$ is
divided by $N$. Bob is confident that because he has kept the decryption
exponent $d$ secret, he and only he possesses the secret key that enables
deciphering of $S.$

At this juncture it is instructive to illustrate the encryption and
decryption of messages in this RSA public key system of Bob's, in particular
the dread message to Bob from Alice that ''The FBI came''. To do so, it
first is necessary to specify the aforementioned symmetric key. A possible
easily usable symmetric key, which Singh suggests$^{22},$ requires Alice to
replace each letter of the alphabet by its ASCII equivalent. ASCII$^{23}$,
the acronym for the American Standard Code for Information Interchange, is
the protocol that converts computer keyboard strokes into the seven-bit
electrical impulses that transmit our email; each such impulse is the binary
representation of a positive integer (less than 2$^{7}=128,$ of course). In
ASCII$^{23}$ the 26 upper case letters A through Z are represented by the
integers (in base 10) 65 through 90$;$ the corresponding lower case letters
are represented by the integers 97 through 122; a space between words is
represented by the integer 64; the ASCII integer representations of other
communication symbols, e.g., the period or the comma, are irrelevant to my
present purpose. Similarly, for my present purpose it is sufficient, as well
as simpler than using ASCII, to ignore the distinction between the upper and
lower cases, and to represent the letters A through Z by the integers 2
through 27 respectively, saving the integer 1 for the space between words.

I therefore take Alice's $C,$ corresponding to her above-quoted dread
message, to be: 21, 9, 6, 1, 7, 3, 10, 1, 4, 2, 14, 6. These integers are
written in base 10 of course, and unless otherwise noted I shall continue to
write integers in the familiar base 10 throughout the remainder of this
paper; the fact that Alice (with Bob's blessing) may choose to write her
integers in another base, or may transmit her $S$ to Bob via email (wherein
the digits 0 through 9 she uses to write her base 10 integers will be
converted into electrical impulses which are the ASCII seven-bit binary
equivalents of the base 10 integers 48 through 57 respectively$^{23}$), in
no way affects the validity of any conclusions drawn below. This point
understood, it next is necessary to choose the pair $p$ and $q$ of primes
whose product yields the $N$ our hypothetical Bob had broadcast for Alice's
use. I will choose the pair 5 and 11, which makes $N$ = 55, a conveniently
small number for my present illustrative purpose but still large enough to
satisfy the requirement that $N$ exceeds every $c$ in $C.$ The quantity $%
\phi $ = 4x10 = 40; so I now can and do choose $e$ = 23$,$ consistent with
the requirement that $e$ be coprime to $\phi .$ To complete Bob's public key
we need a positive integer $d$ such that when $d${\it e }is divided by $\phi 
$ the remainder is unity. The integer $d=$ 7 fits the bill, as the reader
instantly can verify; the convenient method which Bob can use to find $d$ in
actual RSA practice is presented under Subheading IV.D.4.

I now finally am in position to illustrate how Alice constructs her
cryptogram $S$ from $C.$ In order to do so efficiently, however, it is
desirable to introduce the modular arithmetic notation employed in number
theory.$^{24}$

\subsection{Modular Arithmetic Formulation of the RSA Public Key System.}

If the positive integer $b$ is the remainder when the positive integer $a$
is divided by the positive integer $m,$ then $a-b$ is exactly divisible by $%
m.$ In number theory, if a difference of two integers $a$ and $b$ (each of
which now may be positive or negative) is exactly divisible by the positive
integer $m,$ then we say ''{\it a }and $b$ are congruent modulo $m",$ and
write$^{24}$ 
\begin{equation}
a\equiv b\hspace{1in}(%
\mathop{\rm mod}%
\;m).
\end{equation}
Thus the above-specified procedure for obtaining the elements {\it s} of $S$
from the elements $c$ of $C$ can be written as 
\begin{equation}
c^{e}\equiv s\hspace{1in}(%
\mathop{\rm mod}%
\;N).
\end{equation}
Similarly Bob's secret key procedure for deciphering $S$ can be written as $%
c=u,$ where 
\begin{equation}
s^{d}\equiv u\hspace{1in}(%
\mathop{\rm mod}%
\;N).
\end{equation}
Finally, the formula yielding $d$ from $e$ is 
\begin{equation}
de\equiv 1\hspace{1in}(%
\mathop{\rm mod}%
\;\phi ).
\end{equation}

Since Eq. (1) means there is no remainder when $a-b$ is divided by $m,$ Eq.
(1) can be restated as 
\begin{equation}
a-b\equiv 0\hspace{1in}(%
\mathop{\rm mod}%
\;m).
\end{equation}
But if $a-b$ is exactly divisible by $m,$ then so is $a-b-m,$ i.e., if Eq.
(1) holds then it also is true that 
\begin{equation}
a\equiv b+m\hspace{1in}(%
\mathop{\rm mod}%
\;m).
\end{equation}
The pair of Eqs. (1) and (6) imply that Eq. (2), though perfectly correct,
does not uniquely determine $s$ without the additional condition that $s$ is
a positive integer 
\mbox{$<$}%
$N,$ which then guarantees that $s$ indeed is the integer remainder when $%
c^{e}$ is divided by $N.$ Similarly Eq. (3) does not uniquely specify $u$
without the additional condition that $u$ is a positive integer 
\mbox{$<$}%
$N.$

The use of congruences eases Alice's task of constructing her cryptogram $S$
from $C$ via Eq. (2), as Subsection IV.A illustrates; in particular, for the
key number $N$ = 55, encryption exponent $e$ =23, and the $C$ given in the
next to last paragraph of Subsection II.A, Alice's $S$ turns out to be: 21,
14, 51, 1, 13, 27, 10, 1, 9, 8, 49, 51. Using the decryption exponent $d$ =
7 in Eq. (3), Bob then readily decrypts this $S$ into precisely Alice's
original $C,$ as Subsection IV.A also illustrates. The proof that the RSA
system really does enable Bob to correctly decipher every $S$ Alice
transmits, namely (remembering $s$ and $u$ are required to be 
\mbox{$<$}%
$N)$ the proof of the magical fact that Eqs. (2)-(4) imply $u=c$ when $N$ = $%
pq$ and $\phi $ = {\it (p - }1)({\it q - }1), is presented in Subsection
IV.C.

\subsection{Cryptanalysis of RSA System Messages.}

{\it \ }The just displayed illustrative $S$ was obtained from our
illustrative $C$ by successively inserting each individual $c$ into Eq. (2),
i.e. (recalling how Alice constructed her illustrative $C),$ by enciphering
Alice's original plaintext one letter at a time. But insertion of the same $%
c $ into Eq. (2) always yields the same $s;${\it \ }for example since both
the third and last numbers in our illustrative $C$ (recall Subsection II.A)
are 6, both the third and last numbers in our illustrative $S$ turn out to
be 51. In other words our illustrative $S$ is identical to the $S$ into
which Alice's original plaintext would have been enciphered using an
appropriate unique substitution key of the sort described at the outset of
Section II. Of course with actual RSA key numbers $N$ the actually
encountered $s$ in $S$ typically will be very large numbers, not the two
digit numbers 
\mbox{$<$}%
55 of our illustrative $S.$ Nevertheless it now is apparent that, despite
the RSA system's number theoretic sophistication, if Alice continues to
routinely encipher [via Eq. (2)] her plaintext into individual elements $s$
one letter at a time, then the various $S$ she transmits to Bob will be
readily cryptanalysable, without any need to guess Bob's decryption exponent 
$d$ or to employ Eq. (3) at all. More particularly, because Alice does not
attempt to keep secret the messages $S$ she sends to Bob, the relative
frequencies and other characteristics of the various $s$ in her messages,
here assumedly written in English, are readily ascertainable$;$ consequently
the standard deciphering techniques$^{13-15,17}$ applicable to symmetric
unique substitution keys, referred to in the opening paragraphs of Section
II, will be quite employable (e.g., the observation that a relatively
infrequent $s_{1}$ almost invariably is followed by another $s_{2}$ suggests 
$s_{1}$ is $q$ and $s_{2}$ is $u).$

There is no requirement that Alice obtain $S$ by successively inserting each
individual $c$ into Eq. (2) however. For instance Alice simply and
efficiently can obtain a very much more difficult to cryptanalyse $S^{\prime
}$ by first converting her $C$ into a new $C^{\prime }$ composed of integers 
$c^{\prime }$ constructed from large blocks of the original $c$ entries; the
only limit on the sizes of these blocks is that each $c^{\prime }$ must be
less than the key number $N$. Obviously there can be essentially no
available useful information, for the purposes of cryptanalysis, about the
relative frequencies with which blocks of say forty letters occur in
English, especially when during the enciphering those letters can be
interrupted by punctuation marks, as well as by wholly superfluous digit
combinations. Thus, returning now to our illustrative $C,$ whenever Bob's
key number $N$ exceeds (10)$^{40}$ Alice could make her illustrative $S$
very very much harder to decipher by inserting into Eq. (2) not the
individual twelve $c$ comprising the illustrative $C,$ but rather the single
40-digit integer $c^{\prime }=$ 2143790906449201290703821001045802143806.
This $c^{\prime }$ is composed of twenty pairs of digits wherein, reading
from left to right, the pairs whose magnitudes are less than 28 comprise the
original twelve $c$ in their correct order in $C,$ but wherein each pair
exceeding 28 is superfluous and randomly chosen. Alice's insertion of this $%
c^{\prime }$ into Eq. (2) would yield a single $s^{\prime }$ comprising
Alice's entire new sequence $S^{\prime }$ to be transmitted to Bob, which $%
s^{\prime }$ would have no discernible features related to the presence of
those superfluities; yet Bob, after recapturing $c^{\prime }$ from this $%
s^{\prime }$ using Eq. (3), would instantly be able to recognize and discard
the superfluous pairs, i.e., would have no difficulty whatsoever
reconstructing Alice's original ''THE FBI CAME'' from his recaptured $%
c^{\prime }.$ Alice could similarly break up, into successively transmitted
blocks of 40 digits, messages longer than our illustrative ''THE FBI CAME''.

This just described quite simple 40-block scheme is by no means the only
conceivable means of replacing a letter by letter $S$ by an $S^{\prime }$
whose elements $s^{\prime }$ bear no useful relationships (for the purposes
of decipherment via the techniques discussed in the opening paragraphs of
Section II) to the characteristics of the language in which the original
plaintext message was written; moreover, as I will explain in a moment,
modern RSA key numbers $N$ permit blocks considerably larger than merely 40
decimal digits. In short, available perfectly feasible encryption systems
make the likelihood that Alice's RSA-system messages could be deciphered via
the aforementioned techniques, even after receipt by the would-be decipherer
of many openly transmitted messages of hers, virtually zero. On the other
hand those techniques, depending primarily on language properties, are not
the only conceivable means whereby a third party might seek to decipher
Alice's RSA messages to Bob. Descriptions of such deciphering schemes, which
are varied, are beyond the scope of this paper. It is sufficient to state
that no known means of deciphering RSA messages is computationally more
practical than decipherment via factorization of the key number $N.$ In
particular a very thorough highly technical examination$^{25}$ of the
security of the RSA system found (in 1996): ''While it is widely believed
that breaking the RSA encryption scheme is as difficult as factoring the key
number $N,$ no such equivalence has been proven.''; I am not aware of any
later publications which contradict this finding. The relevance of being
able to factor $N$ is that once the primes $p$ and $q$ factoring $N$ are
known, the value of $\phi $ immediately is yielded by the formula $\phi
=(p-1)(q-1)$ [recall Subsection II.A], with which formula, along with Eqs.
(3) and (4), any codebreaker seeking to break Bob's RSA encryption system
presumably will be acquainted; from $\phi $ and Bob's originally publicly
broadcast encryption exponent $e,$ one easily can determine Bob's originally
secret decryption exponent $d$ (as explained under Subheading IV.D.4),
thereby enabling this codebreaker to decipher Alice's cryptogram (which she
did not specially try to keep secret) via Eq. (3) no less readily than Bob
himself can.

\subsection{Modern RSA Systems. Factoring $N=pq$ With Classical Computers.}

It follows that (to the best present understanding) Bob can be confident in
the security of his RSA system, provided there is no practical likelihood
that a would-be codebreaker will be able to deduce the prime factors $p$ and 
$q$ of $N$ from the publicly known quantities $N$ and $e.$ In essence,
therefore, the basis for Bob's confidence is the difficulty of factoring,
with classical computers, numbers $N$ of the astonishingly large magnitudes
typically employed in modern RSA keys. According to recent Internet
publications by RSA Security, Inc., the company founded by the inventors of
the RSA system, key numbers $N$ of 1024 binary digits now are both the
recommended and popularly employed sizes for corporate use$^{26}$; 1024
binary digits corresponds to 309 decimal digits.

The most obvious way to factor a large integer $N$ that is not a prime is to
perform the sequence of divisions of $N$ by the integers 2,3,... $\leq \sqrt{
N}$ until a factor of $N$ is found. If $N$ has only two prime factors, each
of the order of $\sqrt{N},$ then approximately $\sqrt{N}$ divisions will be
required to find the factors of $N.$ Thus, as Ekert$^{27}$ points out, this
straightforward procedure cannot possibly be relied on to yield the prime
factors of really large numbers $N,$ numbers of 100 decimal digits say
(which, though very large by any ordinary standards, are very very much
smaller than the present RSA-recommended key numbers of 309 decimal digits).
For a 100 decimal digit number $N$, i.e., for $N$ of the order of (10)$%
^{100},$ approximately (10)$^{50}$ divisions may be required to ensure
factoring by this straightforward procedure. Even if the average time for a
single division is as small as (10)$^{-12}$ seconds, a very small time
indeed even for today's fastest computers$^{28}$, the total time required to
factor an $N$ $\backsim $ (10)$^{100}$ in this fashion well may be $\backsim 
$ (10)$^{38}$ seconds, a duration very much longer than (6.3)(10)$^{17}$
seconds = 12 billion years, the present best estimate of the age of the
universe.$^{29}$

Nevertheless in 1994 the 129 decimal digits public key number $N$ known as
RSA-129 was factored after only eight months of number crunching, thereby
winning a symbolic \$100 prize Martin Gardner had offered in 1977, shortly
after RSA-129 was first made public.$^{30}$ This accomplishment was made
possible by the ingenuity of mathematicians, who have been able to devise
factoring procedures far more powerful than the brute force procedure
described in the preceding paragraph; in particular RSA-129 was factored
using the so-called {\it quadratic sieve}.$^{31}$ In 1999 RSA-155
(corresponding to a number of 512 binary digits) was factored after no more
than seven months of computing time$^{32}$, using the even more powerful
so-called {\it number field sieve}.$^{33,34}$ This factorization of RSA-155,
in response to the {\it Factoring Challenge}$^{32}$ started in 1991 by RSA
Security, is$^{26}$ the primary reason that RSA Security increased its
recommended key number $N$ size from the previous RSA-155 to the present
RSA-309; indeed RSA Security now recommends a key number size of 2048 binary
digits (i.e., an RSA-617) ''for extremely valuable keys''.$^{26}$ The wisdom
of this recommendation is manifested by two very recent successful
factorizations in response to the Factoring Challenge. Factorization of
RSA-160 (corresponding to a number of 530 binary digits) was announced$^{35}$
on April 1, 2003; the announcement stated that RSA-160 was factored in less
time than RSA-155, and made use of fewer computers in parallel. The
announcement$^{36}$ that RSA-174 (corresponding to a number of 576 binary
digits) had been factored came on December 3, 2003, only eight months later;
as of this writing the time and computer facilities needed to factor RSA-174
have not been released.

That RSA-155 was factored with the expenditure of about the same amount of
computing time as RSA-129 reflects not only the improved power of the number
field sieve over the quadratic sieve, but also the fact that classical
computers had greatly improved in speed during the mere five year interval
from 1994 to 1999. This improvement is expected to continue, as comparison
of the factorization times of RSA-155 and RSA-160 exemplifies. Thus it is
estimated that by 2009 a computer costing no more than \$10 million will be
able to factor RSA-309 in less than one month$^{37};$ correspondingly it is
anticipated that by 2010 the standard (not merely ''extremely valuable'')
RSA key number size will be 2048 binary digits$^{37}$, and by 2030 will be
3072 binary digits$^{37}$ (corresponding to RSA-925)$.$ Inherent in these
anticipations is the well founded belief, thoroughly supported by
experience, that if classical computing is all that's available then RSA
public key systems can be kept secure via increases of key number size no
matter how much classical computers improve, because the magnitude of the
computing effort needed to factor a large number $N$ increases so very
rapidly with increasing $N.$

To be precise, analysis of the number field sieve (presently the most
efficient general-purpose factoring technique$^{34}$ for numbers $N$ of
modern key number size) leads to the conclusion that the number $\nu $ of
bit operations required to factor a large key number $N$ with a classical
computer is expected to increase with $N$ no less rapidly than$^{34,38}$ 
\begin{equation}
\nu (N)=\exp [(1.90)(\ln N)^{1/3}(\ln \ln N)^{2/3}]\cong \exp
[(1.32)L^{1/3}(\log _{2}L)^{2/3}]
\end{equation}
where: $\exp (x)$ denotes the exponential function $e^{x};$ $\ln $ denotes
log$_{e};$ $L=\log _{2}N,$ here and throughout this paper; and a $bit$ $%
operation^{39}$ denotes an elemental computer operation (e.g., the addition
of two bits, each either 0 or 1). The growth of the right side of Eq. (7) as
a function of $L$ is termed$^{40}$ $subexponential,$ i.e., more rapidly than
any power of $L,$ but less rapidly than $\exp (L).$ For a specified
computer, i.e., for some specified number of processors of specified speeds,
the time $\tau (N)$ to perform the factoring should be proportional to $\nu
(N).$ Thus Eq. (7) predicts that the aforementioned hypothesized \$10
million computer, which in 2009 will be able to factor RSA-309 in less than
a month (two weeks say), still will require about sixty million years to
factor RSA-617. We add that even for very large $N$ the computing effort
required to find a pair of primes $p$ and $q$ of magnitude $\sim \sqrt{N}$
is surprisingly small$^{41}$, so that the ability to keep ahead of classical
computer factorization abilities via steadily increasing key number $N$
sizes is not limited by any impracticality in finding key numbers $N$ = $pq.$
Similarly, although it may be thought the increasing encryption and
decryption times inevitably associated with increasing key numbers $N$
ultimately will provide a practical upper bound on the size of usable $N$,
as of the foreseeable future any such bound, though it may exist in
principle, will be utterly inconsequential.$^{42}$

In sum, Bob's confidence in the present and future security of the RSA
systems he now employs and will employ appears to be justified if classical
computing is all that's available. On the other hand, his confidence in the
continued security of his present and future RSA systems would not be well
founded if quantum computers able to employ Shor's algorithm could be
constructed, as this paper now goes on to demonstrate.

\section{Factoring Using Shor's Algorithm.}

Shor's algorithm, which is designed to take advantage of the inherent
potentialities of quantum (as opposed to classical) computers, exploits a
factorization method which differs from the sieves, discussed in Subsection
II.D, that presently are employed for large key number factorization. The
presentation of Shor's algorithm begins, in Subsection III.A immediately
below, with an explanation of this factorization method. The relevant (for a
comprehension of Shor's algorithm) properties of quantum computers are
summarized in Subsection III.B; Subsection III.C, relying on Subsection
III.B, then carefully describes and illustrates Shor's algorithm; some
concluding remarks pertinent to the algorithm are offered in Subsection
III.D. A number of additional remarks about Shor's algorithm are relegated
to a concluding Subsection III.D. Subsections III.A through III.D, taken
together, help clarify the precise import of the assertion that factoring
increasingly large key numbers $N$ = $pq$ ultimately should require less
computational effort, i.e., ultimately should be more feasible, with quantum
computers employing Shor's algorithm than with classical computers.

\subsection{Factoring $N=pq$ Using the Order Property of Integers Coprime to 
$N.$}

Let $n$ denote, here and henceforth unless otherwise stated, a positive
integer coprime to $N$ = $pq$ (meaning, as explained at the outset of
Subsection II.A, that $n$ and $N$ have no common prime factors other than
1), where $p$ and $q$ are two distinct large primes. For any such $n,$ let $%
f_{j},$ $j=1,2,3,...,$ be the remainder when $n^{j}$ is divided by $N.$
Then, as with Eqs. (2) and (3), $f_{j}$ is uniquely specified by the
equation 
\begin{equation}
n^{j}\equiv f_{j}\hspace{1in}(%
\mathop{\rm mod}%
\;pq)
\end{equation}
together with the condition 0 
\mbox{$<$}%
$f_{j}$ 
\mbox{$<$}%
$N.$ As explained in Subsection IV.B, for every $n$ 
\begin{equation}
n^{\phi }=n^{(p-1)(q-1)}\equiv 1\hspace{1in}(%
\mathop{\rm mod}%
\;pq),
\end{equation}
implying $f_{\phi }=1$ for every $n.$ For any given $n,$ however, there well
may exist other integers 1 $\leq $ $j$ $\leq $ $\phi $ $=$ $(p-1)(q-1)$ for
which $f_{j}=1.$ The smallest such $j,$ to be denoted by $r$ hereinafter, is
termed$^{43}$ the $order$ of $n$ modulo $pq.$ Thus, recalling Eqs. (1) and
(5), Eq. (8) for $j=r$ can be restated as 
\begin{equation}
n^{r}-1\equiv 0\hspace{1in}(%
\mathop{\rm mod}%
\;pq).
\end{equation}

Suppose now the order $r$ of some integer $n$ 
\mbox{$<$}%
$N$ and coprime to $N$ is known (how $r$ actually is determined is discussed
below). Suppose further that $r$ is even, necessary in order that $n^{r/2}$
be an integer and thus meaningfully employable in congruences. Then Eq. (10)
implies 
\begin{equation}
(n^{r/2}-1)(n^{r/2}+1)\equiv 0\hspace{1in}(%
\mathop{\rm mod}%
\;pq).
\end{equation}
Because by definition $r$ is the smallest power of $n$ for which Eq. (10)
holds, the factor $(n^{r/2}-1)$ on the left side of Eq. (11) cannot be
exactly divided by $pq,$ i.e., 
\begin{equation}
n^{r/2}-1\not{\equiv}0\hspace{1in}(%
\mathop{\rm mod}%
\;pq)
\end{equation}
The second factor on the left side of Eq.\ (11) is not subject to any such
restriction, i.e., it is possible that 
\begin{equation}
n^{r/2}+1\equiv 0\hspace{1in}(%
\mathop{\rm mod}%
\;pq).
\end{equation}
It is not necessary that Eq. (13) be true, however, i.e., it is perfectly
possible that 
\begin{equation}
n^{r/2}+1\not{\equiv}0\hspace{1in}(%
\mathop{\rm mod}%
\;pq).
\end{equation}
If both Eqs. (12) and (14) hold, we have the situation that the product on
the left side of Eq. (11) is exactly divisible by $pq$ although neither
factor in this product is exactly divisible by $pq.$ It follows that, to
avoid contradiction, one of the factors in the product on the left side of
Eq. (11) [the factor $(n^{r/2}-1)$ say], must be divisible by $p$ but not by 
$q,$ while the other factor [now $(n^{r/2}+1)]$ is divisible by $q$ but not
by $p.$ When the order $r$ of $n$ modulo $N$ is even, therefore, and Eqs.
(12) and (14) both hold, Bob's proclaimed key number $N=pq$ is immediately
factored by computing the following two greatest common divisors (gcd's): of 
$N$ with ($n^{r/2}+1),$ and of $N$ with ($n^{r/2}-1);$ alternatively one can
factor $N$ by first computing $q$ say as the gcd of $N$ with ($n^{r/2}+1),$
and then determining $p$ via division of $N$ by this $q$. The convenient
Euclidean algorithm for finding the gcd of two integers is described under
Subheading IV.D.1.

The probability that a randomly selected $n<N=pq$ and coprime to $N$ will
have an even order $r$ satisfying Eq. (14) is$^{44}$ approximately 1/2.
Moreover, as also is explained under Subheading IV.D.1, calculating the gcd
of a pair of numbers using classical computers, even of a pair of large
numbers like those which are likely to be encountered in the factorization
of an RSA-309 or RSA-617, is a straightforward procedure requiring
negligible computing time (in comparison with the factorization times quoted
in Subsection II.D). Therefore the feasibility of factoring a large $N$ = $%
pq $ via the procedure described in the preceding paragraph depends
primarily on the feasibility of determining the order $r$ of $n$ modulo $N$
for arbitrarily selected $n.$ With classical computers this determination in
essence requires solving the so-called {\it discrete log problem}$^{45};$
experience has shown, however, as Subsection II.D clearly implies and as
Odlyzko$^{46}$ explicitly concludes, that classical factoring of large $N$ = 
{\it pq} via solutions to the discrete log problem is not more feasible than
factoring $N$ using the sieves discussed in Subsection II.D.

With quantum computers on the other hand, determining $r,$ and thereby
factoring $N,$ becomes feasible using the periodicity property of the
sequence $f_{j},$ $j=1,2,3,...,$ defined via Eq. (8). Namely it is proved in
Subsection IV.A that for any $n$ all the integers $%
f_{1},f_{2},...,f_{r-1,}f_{r}=1$ are different, but that for each $j$ in the
range 1 $\leq $ $j$ $\leq $ $r$ and every positive integer $k$ we have $%
f_{j}=$ $f_{j+r}=$ $f_{j+2r}=...$ $f_{j+kr}=$ $f_{j+(k+1)r}=....$ In other
words the sequence $f_{j},$ $j=1,2,3,...,$ is periodic with period $r.$ For
example, returning now to our illustrative $N$ = 55 key number, for $n=16$
the order $r=5$ and the sequence $f_{j}$ is (starting with $j$ = 1)
16,36,26,31,1,16,36,26,31,1,16,36,..., all as is readily verified via the
congruence manipulations discussed and illustrated in Subsection IV.A;
similarly for $n$ = 12 the order $r$ is 4 and the sequence $f_{j}$ is
12,34,23,1,12,34,23,1,12,34,.... Shor's algorithm, taking advantage of the
fact that a quantum computer is described by a wave function (as elaborated
in Subsection III.B), i.e., has wave-like properties, employs the quantum
computer analog of Fourier transformation to extract the order $r$ from a
quantum computer wave function which has been specially constructed to
exhibit this $r$ periodicity for some randomly selected $n.$ Moreover and
most importantly, as also is elaborated below, the computational effort
required to determine $r$ using Shor's algorithm increases with $N$ no more
rapidly than some power of $N,$ i.e. [recall Eq. (7)], increases much more
slowly with $N$ than does the effort required to factor $N$ using a
classical computer. Classical computer factoring via solution of the
discrete log problem does not result in a slower increase of $\nu (N)$ with $%
N$ than Eq. (7) manifests because, $inter$ $alia,$ with such computers
(perhaps since the terms {\it wave function} and {\it wave-like properties}
are meaningless for them), the number of bit operations required to
calculate a Fourier transform is$^{47}$ proportional to $NL$ = $L2^{L},$
i.e., increases with $N$ even more rapidly than does the right side of Eq.
(7).

Before proceeding to the summary, in Subsection III.B, of the relevant
properties of quantum computers, I emphasize (as was foreshadowed in the
closing paragraph of Section I) that once a suitable $r$ has been determined
using Shor's algorithm the factorization of $N$ using Eqs. (12) and (14) can
be routinely performed on a classical computer, making use of the Euclidean
algorithm described under Subheading IV.D.1. Referring to our numerical
illustrations in the preceding paragraph, for the choice $n$ = 12 Shor's
algorithm will yield $r$ = 4. Then from the sequence $f_{j}$ for $n$ = 12 we
need to insert $f_{r/2}=$ $f_{2}$ = 34 [which is congruent to (12)$^{2}$
modulo 55] into Eq. (11). Since Eqs. (12) and (14) both are satisfied for
this $f_{2},$ we immediately know [as explained beneath Eq. (14)], that: (i) 
$f_{2}$ + 1 = 35 must be divisible by one of the factors of 55 (in this case
5, as we would determine by computing the gcd of 35 and 55), and that (ii) $%
f_{2}$ $-$ 1 = 33 must be divisible by the other factor of 55 (in this case
11), as we would determine either by computing the gcd of 33 and 55, or
(more simply) by direct division of 55 by its already determined factor 5.

\subsection{Quantum Computers.}

I assume the readers of this paper have an understanding of quantum
mechanics at the level of standard texts$^{48}$ for introductory quantum
mechanics courses. For such readers, both Mermin$^{1}$ and Grover$^{2}$
intelligibly explain how quantum computers differ from classical computers.
I proceed to very briefly summarize the information about quantum computers
needed to comprehend the functioning of Shor's algorithm, beginning with a
quote from Grover$^{2}$: ''Just as classical computing systems are
synthesized out of two-state systems called bits, quantum computing systems
are synthesized out of two-state systems called $qubits.$ The difference is
that a bit can be in only one of the two states at a time; on the other hand
a qubit can be in both states at the same time.'' To elaborate, any
measurement of the state of a qubit, like any measurement of the state of a
classical bit, can yield only one or the other of two and only two possible
states. Because a qubit is a quantum mechanical system describable by a wave
function, however, the two exclusive possible outcome states for a state
measurement performed on a qubit typically will depend on measurement
details, as of course is not the case for a classical bit. Suppose for
instance that our qubit is a spin 1/2 particle, one of many conceivable
actual physical realizations of a qubit in a practical quantum computer.$%
^{49}$ Then a measurement of the component of the particle's spin along the $%
z$ direction, via a Stern-Gerlach apparatus say, can yield the results +1/2
and -1/2 only, to which correspond respectively orthogonal wave functions
which can be denoted by $|+z\rangle $ and $|-z\rangle $ (using Dirac
notation). Similarly, if a measurement of the component of spin along the $y$
direction is performed on a particle which has been found to have spin +1/2
along the $z$ direction, the only possible results again are +1/2 and -1/2
only, to which correspond respectively orthogonal wave functions which can
be denoted by $|+y\rangle $ and $|-y\rangle .$ But neither of the wave
functions $|+y\rangle $ and $|-y\rangle $ is identical with the wave
function $|+z\rangle $ (or with the wave function $|-z\rangle $ for that
matter). Rather each of the wave functions $|+z\rangle $ and $|-z\rangle $
is a known linear combination of the wave functions $|+y\rangle $ and $%
|-y\rangle $ and vice versa, as readily can be worked out from the
established theory of the behavior of spin 1/2 wave functions under
coordinate axis rotations.$^{50}$

\subsubsection{The Computational Basis. Quantum Computer Wave Functions.}

Therefore, in order to enable convenient employment of a qubit for
computational purposes, namely in order that the two possible outcomes of
state measurements on the qubit be consistently interpretable as
corresponding to the binary integers 0 and 1 respectively (as are
interpretable the outcomes of state measurements on a classical bit), it is
necessary to suppose that the qubit state measurement always will be
performed in the same way, e.g., with a Stern-Gerlach apparatus always lined
up along the positive $z$ direction if the qubit is a spin 1/2 particle. On
this supposition the pair of orthogonal wave functions describing the two
possible qubit state measurement outcomes customarily are denoted by $%
|0\rangle $ and $|1\rangle ;$ these singled-out wave functions comprise the
so-called $computational$ $basis,$ and are interpretable respectively as
corresponding to the binary integers 0 and 1. Thereupon the wave function $%
\Psi $ describing any arbitrary state of the qubit, which always is a linear
superposition of any pair of orthogonal wave functions, typically is
expanded in terms of $|0\rangle $ and $|1\rangle $ only. In other words, for
convenient analysis of the computational functioning of the qubit, one
typically writes 
\begin{equation}
\Psi =\mu |0\rangle +\nu |1\rangle ,
\end{equation}
where $\mu $ and $\nu $ are a pair of complex numbers satisfying

\begin{equation}
|\mu |^{2}+|\nu |^{2}=1.
\end{equation}
Eq. (16), which expresses the fact that $\Psi $ can be and is normalized to
unity (as are $|0\rangle $, $|1\rangle $ and all wave functions discussed
below), permits the interpretation that [when the qubit wave function is
given by Eq. (15)] $|\mu |^{2}$ is the probability that a state measurement
will yield the outcome corresponding to 0, while $|\nu |^{2}$ is the
probability that the same measurement will yield the outcome corresponding
to 1.

A quantum computer is a collection of qubits, and as such also is a quantum
mechanical system whose state must be describable by a normalized wave
function. Consider, in particular, a computer composed of just two qubits,
labeled respectively by $A$ and $B.$ There now are at most 2$\times 2=4$
possible different outcomes of state measurements on the pair of qubits $A,B$
(whether performed simultaneously or successively). Consequently the wave
function $\Psi $ of this computer now must be a linear superposition of at
most four orthogonal two-qubit basis wave functions, which (as Mermin$^{1}$
fully discusses) can be taken to be the computational basis products $%
|0\rangle _{B}|0\rangle _{A}\equiv |00\rangle ,$ $|0\rangle _{B}|1\rangle
_{A}\equiv |01\rangle ,$ $|1\rangle _{B}|0\rangle _{A}\equiv |10\rangle ,$
and $|1\rangle _{B}|1\rangle _{A}\equiv |11\rangle .$ In other words the
most general two-qubit computer wave function has the form 
\begin{equation}
\Psi =\gamma _{00}|00\rangle +\gamma _{01}|01\rangle +\gamma _{10}|10\rangle
+\gamma _{11}|11\rangle ,
\end{equation}
where in the computational basis wave functions $|00\rangle ,$ etc., it is
understood that the two binary digits reading from left to right correspond
to the outcomes of state measurements on qubits $B$ and $A$ respectively,
and where the associated amplitudes $\gamma _{00},$ etc., are complex
numbers satisfying 
\begin{equation}
|\gamma _{00}|^{2}+|\gamma _{01}|^{2}+|\gamma _{10}|^{2}+|\gamma
_{11}|^{2}=1.
\end{equation}
The digit pairs 00, 01, 10 and 11 indexing the computational basis wave
functions appearing in Eq. (17) are the binary system representations of the
integers 0 through 3 (now written in the decimal system), with the proviso
that in the binary system each of these integers is to be represented by no
fewer than two digits. Thus Eq. (17) can be rewritten as 
\begin{equation}
\Psi =\gamma _{0}|0\rangle +\gamma _{1}|1\rangle +\gamma _{2}|2\rangle
+\gamma _{3}|3\rangle
\end{equation}
where the basis wave functions $|i\rangle $ and associated amplitudes $%
\gamma _{i},$ $i=1$ to 3, obviously are merely relabelings, respectively, of
the basis wave functions $|00\rangle ,|01\rangle ,$ etc., and of the
amplitudes $\gamma _{00},\gamma _{01},$ etc.

\subsubsection{Wave function Collapse. The Born Rule.}

In Eqs. (17) and (18) each $|\gamma _{\beta \alpha }|^{2}$ is the
probability that measurements on the qubit pair $A,B$ in the two-qubit state
described by $\Psi $ will yield state $|\alpha \rangle _{A}$ for qubit $A$
and state $|\beta \rangle _{B}$ for qubit $B,$ where each of $\alpha $ and $%
\beta $ can have the values 0 and 1 only, all as developed immediately
above. It is conceivable, however, that the observer will seek to measure
the state of qubit $A$ only, without any attempt to ascertain the state of
qubit $B.$ In this event $|\gamma _{00}|^{2}+|\gamma _{10}|^{2}$ is the
probability of finding $A$ in state $|0\rangle _{A}$, while $|\gamma
_{01}|^{2}+|\gamma _{11}|^{2}$ is the probability of finding $A$ in state $%
|1\rangle _{A}$; we see that these probabilities sum to unity, as they must.
If a measurement on qubit $A$ is performed, and $A$ actually is found to be
in state $|0\rangle _{A}$, then the original wave function $\Psi $ of Eq.
(17) is said to have been $reduced$ or $collapsed$ by the measurement into
the new wavefunction $\Psi ^{\prime }$ = $\Psi _{B}|0\rangle _{A},$ wherein
the one-qubit wave function $\Psi _{B}$ for qubit $B$ is 
\begin{equation}
\Psi _{B}=[|\gamma _{00}|^{2}+|\gamma _{10}|^{2}]^{-1/2}[\gamma
_{00}|0\rangle _{B}+\gamma _{10}|1\rangle _{B}].
\end{equation}
Eq. (20) is in accordance with the $Born$ $rule,$ which Mermin$^{1}$ fully
discusses; the normalizing factor $[|\gamma _{00}|^{2}+|\gamma
_{10}|^{2}]^{-1/2}$ in Eq. (20) is needed to ensure that $\Psi ^{\prime }$
and $\Psi _{B}$ are normalized wave functions, i.e., that in $\Psi ^{\prime
} $ and in the state of $B$ described by the one-qubit wave function $\Psi
_{B},$ the individual probabilities of finding qubit $B$ in state $|0\rangle
_{B}$ and in state $|1\rangle _{B}$ sum to unity, as again they must. Note
that the square of the coefficient of $|\beta \rangle _{B}$ in Eq. (20),
which represents the probability of finding qubit $B$ in the state $|\beta
\rangle _{B}$ {\it knowing} that a measurement on qubit $A$ in the two-qubit
state described by $\Psi $ of Eq. (17) already has yielded $|0\rangle _{A},$
differs from $|\gamma _{\beta 0}|^{2}$ representing the probability, {\it %
without any such knowledge}, that measurements on the qubit pair $A,B$ in
the two-qubit state described by $\Psi $ will find qubit $A$ in state $%
|0\rangle _{A}$ and qubit $B$ in state $|\beta \rangle _{B}$. The
modification of Eq. (20) appropriate to the circumstance that $A$ actually
had been found in state $|1\rangle _{A}$ rather than in state $|0\rangle
_{A} $ is obvious. Equally obviously [starting again with the two-qubit
system in the state described by $\Psi $ of Eq. (17)]: (i) the probability
of finding qubit $B$ in state $|\beta \rangle _{B}$ (here $\beta $ is either
0 or 1) without any attempt to ascertain the state of qubit $A$ is $%
\sum_{\alpha }|\gamma _{\beta \alpha }|^{2},$ with the sum of these
probabilities = $\sum_{\beta }\sum_{\alpha }|\gamma _{\beta \alpha }|^{2}=1;$
and (ii) if $B$ actually is found in state $|\beta \rangle _{B}$ ($\beta $
again either 0 or 1), the original $\Psi $ collapses into the wave function $%
\Psi _{A}|\beta \rangle _{B},$ where 
\begin{equation}
\Psi _{A}=[|\gamma _{\beta 0}|^{2}+|\gamma _{\beta 1}|^{2}]^{-1/2}[\gamma
_{\beta 0}|0\rangle _{A}+\gamma _{\beta 1}|1\rangle _{A}].
\end{equation}

The considerations in the two preceding paragraphs are immediately
extensible to larger quantum computers, composed of $g$ $\geq 2$ qubits.
Since a state measurement on any given qubit can have at most two different
outcomes (which we have designated by the binary system digits 0 and 1),
state measurements on the entire collection of qubits comprising a $g$-qubit
quantum computer can have at most 2$^{g}$ different outcomes.
Correspondingly the wave function $\Psi $ describing any state of a $g$
-qubit quantum computer is a linear superposition of at most 2$^{g}$
orthogonal $g$-qubit basis wave functions. Index these $g$ qubits by $k$
running from 1 to $g.$ Then the 2$^{g}$ computational basis wave functions
for the computer can be taken to be 
\begin{equation}
|0\rangle _{g}|0\rangle _{g-1}...|0\rangle _{2}|0\rangle _{1}\equiv
|00...00\rangle ,\quad |0\rangle _{g}|0\rangle _{g-1}...|0\rangle
_{2}|1\rangle _{1}\equiv |00...01\rangle ,...|1\rangle _{g}|0\rangle
_{g-1}...|0\rangle _{2}|1\rangle _{1}\equiv |10...01\rangle ,
\end{equation}
etc., and [analoguously to Eq. (19)] the most general $g$-qubit quantum
computer wave function can be expressed as 
\begin{equation}
\Psi =\sum_{i=0}^{2^{g}-1}\gamma _{i}|i\rangle ,
\end{equation}
with of course 
\begin{equation}
\sum_{i=0}^{2^{g}-1}|\gamma _{i}|^{2}=1.
\end{equation}
In Eqs.(23) and (24): the integers $i$ are conveniently written in the
decimal system, as in Eq. (19); the binary system representation of each $i$
consists of no fewer than $g$ digits; each computational basis wave function 
$|i\rangle $ represents a $g$-qubit state wherein (for every $k,$ with $%
1\leq k\leq g)$ the outcome (either 0 or 1) of a state measurement on the k%
{\it th} qubit surely equals the k{\it th} digit (reading from right to
left) in the binary system representation of $i;$ and $|\gamma _{i}|^{2}$ is
the probability that when the computer is in the state described by the wave
function $\Psi $ of Eq. (23), state measurements on the collection of $g$
qubits will have the characteristic outcomes (specified earlier in this
sentence) of state measurements when the computer wave function is simply $%
|i\rangle $. Moreover if, while the computer is in the state described by $%
\Psi ,of$ Eq. (23), the computer operator were to measure, e.g., the states
of qubits 1, 2 and $g,$ and were to obtain the outcomes $|1\rangle _{1},$ $%
|0\rangle _{2}$ and $|1\rangle _{g}$ respectively, these measurements will
collapse $\Psi $ into the wave function $\Psi _{g-3}[|1\rangle _{1}|0\rangle
_{2}|1\rangle _{g}],$ where: (i) the ({\it g-3)-}qubit wave function 
\begin{equation}
\Psi _{g-3}=[\sum_{j}|\gamma _{j}|^{2}]^{-1/2}\sum_{j}\gamma _{j}|j\rangle
\end{equation}
describes the state of the remaining qubits 3,4,...( g-1) knowing that state
measurements on qubits 1, 2 and $g$ in the $g$-qubit system described by $%
\Psi $ had yielded the outcomes $|1\rangle _{1},$ $|0\rangle _{2}$ and $%
|1\rangle _{g}$ respectively; and (ii) in Eq. (25) $j$ runs over all
integers whose $g$-digit binary representations begin with 1 and end with 01
(now reading from left to right).

\subsubsection{Operations on Quantum Computers. Unitarity.}

It can be assumed that the $g$-qubit quantum computer wave function $\Psi $
of Eq. (23), like the wave function of any other quantum mechanical system,
evolves in time in accordance with the non-relativistic time-dependent
Schroedinger equation 
\begin{equation}
\frac{\partial \Psi }{\partial t}=\frac{ih}{2\pi }{\bf H}\Psi ,
\end{equation}
where $h$ is Planck's constant and the Hamiltonian {\bf H}, which{\bf \ }may
be time dependent, is some appropriate Hermitian operator capable of
meaningfully acting on the various computational basis wave functions $%
|i\rangle $ appearing in Eq. (23). In this circumstance the wave function $%
\Psi (t)$ at any time $t$ $\geq $ 0 is related to the wave function $\Psi
(0) $ at $t=0$ by 
\begin{equation}
\Psi (t)={\bf U}\Psi (0),
\end{equation}
where, because ${\bf H}$ is Hermitian, ${\bf U\equiv U}(t)$ is a linear
normalization-conserving operator$^{51},$ i.e., a $unitary$ operator.$^{52}$
Whatever the physical realizations of the individual qubits comprising the
quantum computer may be, the computer's utility as a computational tool
depends on the ability (of the person performing the computation) to control
the time evolution of its wave function. But this desired controlled
evolution, which generally requires modifying the environments of the
individual qubits (e.g., when the qubits are spin 1/2 particles, rotating
the individual magnetic fields acting on those particles), still necessarily
is a time evolution of $\Psi $ under Eq. (26). Thus the desired controlled
evolution also is described by Eq. (27), i.e., necessarily involves a
unitary operation on the initial wave function $\Psi (0),$ here the computer
wave function when the modifications of the individual qubit environments
were initiated.

Accordingly each planned operation in the sequence of operations
constituting any proposed quantum computing algorithm, e.g., Shor's
algorithm, must be a unitary operation. Postulated non-unitary operations on
a quantum computer, no matter how attractive seemingly, are irrelevant and
thus of no interest whatsoever to the use of the computer as a computational
tool, because no non-unitary operation will be attainable with any actual
physical realizations of the qubits comprising the computer. Therefore it is
important to note (as we amplify in various Subheadings under Subsection
III.C below) that each of the quantum computing operations involved in
Shor's algorithm indeed is unitary. Of course the impossibility of
constructing a physical realization of any non-unitary operation does not
imply every conceivable unitary operation on a quantum computer can be
physically realized; furthermore if the computer is composed of a large
number of qubits, e.g., thousands of qubits (as is quite likely in actual
practical applications of Shor's algorithm, see under Subheading III.C.1
below), the prospect of actually constructing a physical realization of any
non-trivial unitary operation ${\bf U}$ on so large a collection of qubits
seems hopeless at first sight, even if there is reason to believe that a
physical realization of ${\bf U}$ must exist. Fortunately, however, and
absolutely crucial for the practical potential of quantum computation, it
can be proved that every conceivable unitary operation on an arbitrarily
large $g$-qubit quantum computer, even an operation involving simultaneous
modifications of the environments of all $g$ $>>2$ qubits, can be reproduced
via an appropriate sequence of basic unitary one-qubit and two-qubit
operations only.$^{53}$ Moreover, and also very worthy of note, numerous
methods, based solely on known physics, for achieving physical realizations
of these basic unitary one-qubit and two-qubit operations (also known as $%
universal$ $quantum$ $gates^{53})$ have been proposed$^{49}$, although
admittedly the actual implementation of many of these conceptual physical
realizations well may prove to be difficult in practice.

For the purposes of this paper, therefore, it is not unreasonable to assume
that quantum computers consisting of arbitrarily large assemblages of
qubits, capable of performing any desired computational algorithm which can
be formulated in terms of unitary operations, eventually will be
constructed. Granted this assumption, a measure of the quantum computational
effort required to perform any given algorithm, indeed the only obvious
measure, is the number of universal quantum gates that must be strung
together to perform the algorithm on a quantum computer; in essence the
universal quantum gate operations play the role, for quantum computation,
that the bit operations referred to in connection with Eq. (7) play for
classical computation. Thus we now finally are able to make precise, as we
very much need to do, the meaning of our oft-repeated assertion that the
Shor algorithm enables a quantum computer to factor large key numbers $N$ = $%
pq$ with far less computational effort than using a classical computer
requires. In particular, with a quantum computer using Shor's algorithm the
number $\nu _{q}$ of universal quantum gates required to determine an order $%
r$ that will enable factorization of a large $N$ = $pq$ via Eq. (11) is
estimated$^{3,54}$ to obey the equation 
\begin{equation}
\nu _{q}(N)=O{\bf [(}\ln N)^{2}(\ln \ln N)(\ln \ln \ln N)]=O[L^{2}(\log
_{2}L)(\log _{2}\log _{2}L)],
\end{equation}
where $L$ = $\log _{2}N$ as in Eq. (7) and the symbol {\it O}, denoting {\it %
Order of}, means$^{55}$ there exists a constant $K$ such that for
sufficiently large $N$%
\begin{equation}
\nu _{q}(N)\leq K[L^{2}(\log _{2}L)(\log _{2}\log _{2}L)].
\end{equation}
In connection with Eqs. (28) and (29) it is useful to recognize that for
large $N$ the number of qubits required to represent $N$ is essentially $L.$
To be precise, for any real number $x\geq 1,$ let [$x]$ denote the largest
integer $\leq $ $x;$ then it is easily seen that the number of qubits needed
to represent any $N$ is [$\log _{2}N]$ +1, which for large $N$ differs
negligibly from $L.$

The immediately preceding discussion has overlooked the fact that in
practice the actual factorization of $N$ using Shor's algorithm requires
computational operations (e.g., classical computer gcd calculations as
discussed at the end of section III.A) beyond the universal quantum gate
operations whose number is estimated in Eq. (28). Subsection III.D below
implies, however, that for the purposes of this paper such neglected
computational operations do not negate the validity of Eq. (29) as a measure
of the computational effort required to factor a large $N$ = $pq$ with a
quantum computer using Shor's algorithm. Therefore comparison of Eqs. (7)
and (29) validly quantifies the reduction in the computational effort
required to factor a large $N$ that is achievable with a quantum computer.
Whereas according to Eq. (7) the number of elemental computer operations
needed to accomplish the factorization of $N$ with a classical computer
increases faster than any power of $L$ = $\log _{2}N,$ the needed number of
elemental computer operations using a quantum computer increases only barely
more rapidly than $L^{2}$ (indeed surely less rapidly than $L^{3})$
according to Eq. (29). To illustrate the practical import of this reduction,
let us repeat the numerical exercise presented immediately below Eq. (7),
only this time for a quantum computer. For any specified sufficiently large
quantum computer (i.e., for any quantum computer composed of sufficiently
many qubits to handle the Shor algorithm determination of $r$ for all
relevant $N,$ see under Subheading III.C.1), the time $\tau _{q}(N)$ needed
to complete the factorization of a large $N$ should be approximately
proportional to $\nu _{q}(N)$ given by Eq. (29), irrespective of the value
of $K$ appropriate to that computer. Suppose we were able to construct a
quantum computer which, like the classical computer we previously
hypothesized, could factor RSA-309 in two weeks time. Then this same quantum
computer (again assumedly composed of sufficiently many qubits) should be
able to factor RSA-617 in no more than about 9 weeks, in contrast to the 60
million years the classical computer which factored RSA-309 would require.

Moreover it is not unreasonable to believe that a sufficiently large quantum
computer, if such computers can be built at all, will be able to factor
RSA-309 in two weeks. Two weeks is about 1.2$\times 10^{6}$ seconds. For
RSA-309, i.e., for $L$ = 1024, the value of $\nu _{q}$ from Eq. (29) is only
3.5$\times 10^{9}$ even assuming $K$ will be as large as 100, which seems
doubtful. To factor RSA-309 in two weeks, therefore, the average time for
performing a quantum gate operation need be no faster than about 300
microseconds, which should be no problem whatsoever for quantum computer
elements, whether operating on atomic, molecular or photonic scales. In
short, once sufficiently large quantum computers become available Bob no
longer will be justifiably confident that, merely via increases of his
proclaimed key number size, he can maintain the security of Alice's
RSA-coded messages to him in the face of anticipated likely improvements in
quantum computer capabilities.

Before closing this discussion of operations on quantum computers it is
important to note that wave function collapsing measurements on any part of
a quantum computer, though normalization conserving by virtue of the Born
rule$^{1}$ referred to under Subheading III.B.2, are not--strictly
speaking--quantum computing operations of the sort discussed in earlier
paragraphs under the present Subheading. In particular, let $\Psi _{M}$
denote the wave function describing the state of the $g-G$ remaining qubits
after observing the outcomes of state measurements on any {\it G}-qubit
subset (1$\leq G\leq g)$ of a $g$-qubit quantum computer described by the
wave function $\Psi $ of Eq. (23); for instance $\Psi _{M}$ might be the
wave function $\Psi _{g-3}[|1\rangle _{1}|0\rangle _{2}|1\rangle _{g}]$
introduced in connection with Eq. (25). Then, as Mermin$^{1}$ discusses,
because both $\Psi _{M}$ and $\Psi $ are normalized wave functions
expressible as linear superpositions of the very same set of 2$^{g}$
orthogonal computational basis wave functions, there must exist a unitary
operator {\bf U}$_{M}$ such that 
\begin{equation}
\Psi _{M}={\bf U}_{M}\Psi .
\end{equation}
Since $\Psi $ can be thought of as $\Psi (0),$ the computer wave function at
time $t=0$ when the measurement operation began, and $\Psi _{M}$ can be
thought of as $\Psi $($t),$ the computer wave function at time $t>0$ when
the measurement has been completed, Eq. (30) appears to be of precisely the
same form as Eq. (27). The subtle difference is that whereas in Eq. (27) we
have been considering unitary operators which are predictably controllable
[i.e., which during each step of the computational algorithm will cause the
computer wave function $\Psi (0)$ to evolve into some desired $\Psi $($t)],$ 
${\bf U}_{M}$ of Eq. (30) generally is not predictably controllable. Rather
the ${\bf U}_{M}$ we obtain as a result of the measurement generally is only
one of many possible ${\bf U}_{M},$ whose likelihoods of turning up in the
actual measurement operation we have performed depend on the values of the
coefficients $\gamma _{i}$ in Eq. (23); only after we observe the
measurement outcome can we decide which of the many possible ${\bf U}_{M}$
actually has been obtained. Correspondingly, there generally is no way,
before the measurement operation, to introduce a sequence of universal
quantum gates which will reproduce the unitary operator ${\bf U}_{M}$ of Eq.
(30) that actually is attained.

\subsection{The Operations Constituting Shor's Algorithm.}

Shor's original formulation$^{3}$ of his algorithm has been given an
admirably readable (by non-specialists) step-by-step prescription by
Williams and Clearwater$^{56},$ which my presentation will closely follow,
but also will expand on and illustrate. The text of each subheading in this
Subsection very briefly describes one of those eight steps.

\subsubsection{Determine the Minimum Computer Size Required. Divide the
Qubits into Two Registers.}

The reader is reminded of the notation and contents of Subsection III.A.
Shor's algorithm seeks to accurately discern the periodicity with period $r$
manifested by the sequence $f_{j},$ $j=1,2,3,...,$ obtained from Eq. (8) for
some chosen $n.$ In order to do so, the algorithm must operate on sequences
which are many many periods in length (see under Subheading III.C.7), much
as in conventional classical Fourier transformation. In actual practice the
order $r$ well may attain its maximum possible value ($p-1)(q-1)/2$, which
for large $N$ is likely to be only very slightly smaller than half the
to-be-factored $N=pq$ (see Subsection IV.B)$;$ for instance in the case of
our illustrative $N$ = 55, the order $r$ equals its maximum allowed value 20
for fully 16 of the 40 integers $n<55$ that are coprime to $55,$ including $%
n $ as small as 2 and as large as 53. Consequently accurate determination of 
$r $ using Shor's algorithm generally requires the use of powers $j$ 
\mbox{$>$}%
\mbox{$>$}%
$N$ in Eq. (8). Shor$^{3}$ recommends (in effect) that the maximum power $%
j=j_{\max }$ employed be no less than $N^{2},$ a recommendation this paper
accepts; Williams and Clearwater$^{56}$ recommend $j_{\max }$ be even
greater, namely at least 2$N^{2}.$ Thus, following Shor, the quantum
computer being employed to determine $r$ via Shor's algorithm must contain
at least enough qubits to represent powers $j$ up to $j_{\max }$ = $N^{2}.$
The minimum number of qubits needed to represent the integer $N^{2}$ is [$%
\log _{2}N^{2}]$ +1 [recall the text immediately following Eq. (29)]$;$
accordingly, the quantum computer will be supposed to contain a set of $y$ =
[log$_{2}(N^{2}]$ + 1 qubits, which will be said to comprise $register$ {\it %
Y}. In addition the computer must contain a second set of qubits, here
termed $register$ {\it Z}, capable of storing the computed values of $f_{j},$
which can be as large as $N-1;$ the size of this register will be taken to
be its minimum possible value $z=$ [$\log _{2}(N-1)]$ +1 qubits. Note that
because $N$ is known to be the product of a pair of odd primes, i.e., surely
is not a power of two, 2$^{y-1}<N^{2}<2^{y}.$

With the large $N$ of interest herein, the difference between [$\log
_{2}(N^{2})]+1$ and 2$L$, like the difference between [$\log _{2}(N-1)]$ and 
$L$ or between $L$ and $L+1,$ is negligible for the purpose of estimating
the computational efforts required to accomplish the various individual
steps constituting Shor's algorithm; $L$ = $\log _{2}N,$ as previously. Thus
in any subsequent estimates herein of those efforts $y$ justifiably will be
replaced by 2$L$ (ignoring the fact that 2$L$ need not be an integer); this
is the same replacement for $y$ employed$^{54}$ to obtain Eq. (28). For the
purpose of such estimates the difference between [$\log _{2}(N^{2})]$ $\cong
2L$ and [[$\log _{2}(2N^{2})]$ $\cong $ $2L+1$ also is negligible, meaning
that the estimated computational efforts required to accomplish the various
individual steps constituting Shor's algorithm actually do not depend
significantly on whether we prefer the Shor or the Williams-Clearwater
estimates of the required $j_{\max }.$ Furthermore we now can conclude that
unless for large $N$ the actually required value of $j_{\max }$ differs from
Shor's recommended value by quantities exceeding {\it O(N}$^{2})$ [where 
{\it O }is defined as in Eq. (28)], determining $r$ and thereby factoring a
large $N$ will require a quantum computer not less than about 3$L$ qubits in
size. In other words, factoring a key number of the presently recommended
size RSA-309 corresponding to 1024 binary digits (recall Subsection II.D)
seemingly would require a quantum computer of at least 3072 qubits in size;
factoring RSA-617 would require a quantum computer of more than 6,000 qubits
in size.

$.$

\subsubsection{Load the First Register With the Integers Less Than or Equal
to $2^{y}-1.$}

After ordering and indexing the $y$ qubits in register $Y$ as discussed in
connection with Eqs. (22)-(25), the complete set of computational basis wave
functions for those qubits can be written as $|j\rangle _{Y},$ where: the
subscript indicates that we are writing wave functions for register $Y;$ $j$
is an integer, 0 $\leq $ $j\leq $ 2$^{y}-1,$ which shall be written in
decimal notation; when register $Y$ is in the state described by the basis
wave function $|j\rangle _{Y},$ the binary digit representation of $j$
immediately reveals the one-qubit basis state, $|0\rangle _{k}$ or $%
|1\rangle _{k},$ of each of the qubits in register $Y$; and it is understood
that qubit $k$ (1 $\leq $ $k$ $\leq $ $y),$ whose basis states are
identified by the subscript $k,$ corresponds to the $k$th digit, reading
from right to left, in the binary system representation of $j$ [as discussed
immediately beneath Eq. (24)]. The computational basis wave functions for
register $Z$ similarly are denoted by $|i\rangle _{Z},$ where 0 $\leq $ $%
i\leq $ 2$^{z}-1.$ It is postulated that initially every one of the $y+z$
qubits constituting the quantum computer can be set into its own one-qubit $%
|0\rangle $ basis state, i.e., that the initial wave function of the entire
quantum computer is $\Psi _{C}^{(0)}$ = $|0\rangle _{Y}|0\rangle _{Z};$ here
and hereinafter, the subscript $C$ denotes the wave function of the entire
computer. Proceeding with the algorithm requires transforming the initial
register $Y$ wave function $\Psi _{Y}^{(0)}\equiv $ $|0\rangle _{Y}$ to its
first stage form 
\begin{equation}
\Psi _{Y}^{1S}=2^{-y/2}\sum_{j=0}^{2^{y}-1}|j\rangle _{Y},
\end{equation}
i.e., requires replacing the initial $|0\rangle _{Y}$ [wherein a measurement
of the state of register $Y$ can only yield the result 0] by the sum on the
right side of Eq. (31) [wherein, recalling the discussion of Eqs. (23)-(24),
a measurement of the state of register $Y$ has an equal chance of yielding
any of the integers between 0 and 2$^{y}-1$ inclusive]. There are 2$^{y}$
independent $|j\rangle _{Y}$ on the right side of Eq. (31); thus the factor $%
2^{-y/2}$ guarantees $\Psi _{Y}^{1S}$ is normalized.. Since we know $%
N^{2}\leq $ 2$^{y}-1,$ the sum in Eq. (31) includes every $j$ less than or
equal to Shor's recommended $j_{\max }=N^{2}.$

The transformation of $|0\rangle _{Y}$ to $\Psi _{Y}^{1S}$ of Eq. (31) is
accomplished via use of the one-qubit operation {\bf U}$_{H}$ known as a 
{\it Hadamard} transformation, which is defined$^{57}$ so that the results
of the Hadamard operation on the one-qubit basis state wave functions $%
|0\rangle $ and $|1\rangle $ are 
\begin{equation}
{\bf U}_{H}|0\rangle =\frac{1}{\sqrt{2}}(|0\rangle +|1\rangle ),\;\hspace{%
0.5in}{\bf U}_{H}|1\rangle =\frac{1}{\sqrt{2}}(|0\rangle -|1\rangle ).\;%
\hspace{1in}
\end{equation}
{\bf U}$_{H}$ is known$^{57}$ to be unitary, as is trivially verifiable; the
factor 1/$\sqrt{2}$ in Eq. (32) enables {\bf U}$_{H}$ to preserve
normalization, as we know a unitary operation must.$^{52}$ Denote the
operation that performs the Hadamard on qubit $k$ alone, without affecting
any other qubits, by {\bf U}$_{Hk}.$ Next consider the result of operating
with {\bf U}$_{Hk}$ on a computational basis wave function $|j\rangle _{Y}$
for which the $kth$ digit (reading from right to left) in the binary
expansion of $j$ is zero ($not$ 1), meaning [recall Eqs. (22)-(25)] that the
product of one qubit basis wave functions constituting $|j\rangle _{Y}$
includes the factor $|0\rangle _{k}$ ($not$ $|1\rangle _{k}$). To obtain
this desired result, we need only put the subscript $k$ on every one of the
basis wave functions in Eq. (32); moreover since our present $|j\rangle _{Y}$
contains no $|1\rangle _{k}$ basis state, we are here concerned only with
the first equality in Eq. (32). It follows that, except for the factor 1/$%
\sqrt{2},$ the operation {\bf U}$_{Hk}$ on our present $|j\rangle _{Y}$
merely replaces $|0\rangle _{k}$ in $|j\rangle _{Y}$ by $|0\rangle
_{k}+|1\rangle _{k}$ while leaving $|j\rangle _{Y}$ otherwise unchanged. In
the binary expansion of the integer $j,$ however, changing the $k$th digit
from 0 to 1 (always reading from right to left) produces the binary
expansion of the integer $j+2^{k-1}.$ Therefore when $|j\rangle _{Y}$
contains no $|1\rangle _{k}$ basis state, 
\begin{equation}
{\bf U}_{Hk}|j\rangle _{Y}=\frac{1}{\sqrt{2}}\{|j\rangle
_{Y}+|j+2^{k-1}\rangle _{Y}\}.
\end{equation}

Now perform the $y$ operations ${\bf U}_{H1},{\bf U}_{H2},{\bf U}_{H3},...,%
{\bf U}_{Hy},$ sequentially (first ${\bf U}_{H1},$ second ${\bf U}_{H2},$
etc., ) on the initial register $Y$ wave function $|0\rangle _{Y}\equiv \Psi
_{Y}^{(0)}.$ We know $|0\rangle _{Y}$ contains no $|1\rangle _{i}$ factors
for any $i,$ $1\leq $ $i$ $\leq $ $y.$ Thus we surely can employ Eq. (33)
for the first of these operations to obtain 
\begin{equation}
\Psi _{Y}^{(1)}={\bf U}_{H1}|0\rangle =\frac{1}{\sqrt{2}}\{|0\rangle
_{Y}+|0+2^{0}\rangle _{Y}\}=\frac{1}{\sqrt{2}}\{|0\rangle _{Y}+|1\rangle
_{Y}\}=\frac{1}{\sqrt{2}}\sum_{j=0}^{1}|j\rangle _{Y}.
\end{equation}
Because ${\bf U}_{H1}$has been defined so that it performs the Hadamard
operation on qubit 1 only, the wave function $\Psi _{Y}^{(1)}$ (like $\Psi
_{Y}^{(0)})$ does not contain the factor $|1\rangle _{2},$ as is directly
evidenced by the fact that both the integers 0 and 1 on the right side of
Eq. (34) are less than 2. Consequently we also can employ Eq. (33) for the
second of these sequential operations, thereby finding for ${\bf U}_{H2}{\bf %
U}_{H1}|0\rangle =$ ${\bf U}_{H2}\Psi _{Y}^{(1)}\equiv \Psi _{Y}^{(2)},$

\begin{equation}
\Psi _{Y}^{(2)}=\frac{1}{\sqrt{2}}{\bf U}_{H2}\{|0\rangle _{Y}+|1\rangle
_{Y}\}=\frac{1}{2}\{[|0\rangle _{Y}+|0+2\rangle _{Y}]+[|1\rangle
_{Y}+|1+2\rangle _{Y}]\}=\frac{1}{2}\sum_{j=0}^{3}|j\rangle _{Y}.
\end{equation}
Because every one of the integers on the right side of Eq. (35) is less than
4 = 2$^{2},$ $\Psi _{Y}^{(2)}$ surely does not contain the factor $|1\rangle
_{3},$ permitting use of Eq. (33) to evaluate ${\bf U}_{H3}\Psi _{Y}^{(2)}.$
Proceeding in this fashion, it is readily seen that the result of the full
sequence of Hadamard operations on $|0\rangle _{Y}$ is 
\begin{equation}
\Psi _{Y}^{(y)}={\bf U}_{Hy}{\bf U}_{H(y-1)}...{\bf U}_{H2}{\bf U}%
_{H1}|0\rangle =\left( \frac{1}{\sqrt{2}}\right)
^{y}\sum_{j=0}^{2^{y}-1}|j\rangle _{Y}.
\end{equation}

The right side of Eq. (36) is the desired $\Psi _{Y}^{1S}$ of Eq. (31). It
is generally agreed$^{58}$ that the above-defined Hadamard one-qubit
operations are universal quantum gates, as defined under Subheading III.B.3.
Accordingly accomplishing this first stage transformation of the initial $%
|0\rangle _{Y}$ to $\Psi _{Y}^{1S}$ requires no more than $y=$ 2$L$
universal quantum gates.

\subsubsection{Select an $n.$ For Each $j$ in register $Y,$ Place the
Remainder $f_{j}$ $\equiv $ $n^{j}$ (Modulo $N)$ Into Register $Z.$}

After this just completed first stage, the wave function of the entire
computer is $\Psi _{C}^{1S}$ = $\Psi _{Y}^{1S}|0\rangle _{Z},$ meaning that
after competion of the first stage a state measurement on register $Z$ still
is guaranteed to yield the integer 0 only, irrespective of what $j$ may be
yielded by a simultaneous state measurement on register $Y.$ For the next
step an $n$ coprime to $N$ is required. As Subsection IV.B explains, such an 
$n$ can be obtained, with probability essentially indistinguishable from
unity, simply by choosing an arbitrary integer $i,$ either in the range 1 
\mbox{$<$}%
$i<$ $N$ or in the range 1 
\mbox{$<$}%
$i<$ $2^{y}-1.$ Of course whether any selected integer $i$ actually is
coprime to $N$ readily can be tested by calculating the gcd of $i$ and $N$
using a classical computer (as discussed under Subheading IV.D.1), but the
probability such a randomly chosen $i$ will not be coprime to $N$ is so
small the effort of computing this gcd does not seem worthwhile. If the
selected $i$ is not coprime to $N,$ that fact will become apparent when it
is realized the value of the supposed order $r,$ inferred as in step 7
below, does not satisfy Eq. (10); Subsection IV.B explains that no integer $%
r $ can satisfy Eq. (10) when the integer $n$ in Eq. (10) actually is not
coprime to $N.$ In this event it merely will be necessary to repeat steps 2
through 7 after choosing a different $i,$ which new $i$ essentially
certainly will be coprime to $N;$ such repetitions often are required even
when the chosen $i$ indeed is coprime to $N$ (see under Subheading III.C.8
below).

Assuming now an $i\equiv n$ coprime to $N$ actually has been selected, the
next step of the algorithm transforms $\Psi _{C}^{1S}$ to its second stage
form 
\begin{equation}
\Psi _{C}^{2S}=\left( \frac{1}{\sqrt{2}}\right)
^{y}\sum_{j=0}^{2^{y}-1}|j\rangle _{Y}|f_{j}\rangle _{Z},
\end{equation}
where $f_{j}$ is defined by Eq. (8). With the computer wave function $\Psi
_{C}^{2S}$ of Eq. (37), the result of a state measurement on the collection
of qubits in register $Z$ must yield one of the remainder integers 1$\leq $ $%
f_{j}$ $\leq $ $N-1$ prescribed by Eq. (8). Moreover, because of the
periodicity of $f_{j}$ demonstrated in Subsection IV.A, every one of the $%
f_{j}$ in Eq. (37) must equal one of the (all necessarily different) $f_{1},$
$f_{2},...,$ $f_{r}=f_{0}=1.$ No other integers can result from a state
measurement on register $Z$ after completion of the second stage of the
algorithm; in particular, since $n$ is coprime to $N$ by definition, such a
measurement now cannot possibly yield the previously (at completion of the
first stage) assured result 0.

I shall not detail the operation which transforms $\Psi _{C}^{1S}$ to $\Psi
_{C}^{2S}$ of Eq. (37). The operation is fully discussed by Shor$^{3},$ who
shows it is unitary. The number of universal quantum gates required to
perform this unitary operation is$^{3,54}$ $O[L^{2}(\log _{2}L)(\log
_{2}\log _{2}L)]$. Let us illustrate Eq. (37) when $N$ = 55 and $n$ =16. In
these circumstances, as discussed in Subsection III.A, $r$ = 5 and the
sequence $f_{j}$ (now starting with $j$ = 0) is
1,16,36,26,31,1,16,36,26,31,1,16,36,.... Accordingly in this illustrative
case Eq. (37) is 
\[
\Psi _{C}^{2S}=2^{-y/2}\{|0\rangle _{Y}|1\rangle _{Z}+|1\rangle
_{Y}|16\rangle _{Z}+|2\rangle _{Y}|36\rangle _{Z}+|3\rangle _{Y}|26\rangle
_{Z}+|4\rangle _{Y}|31\rangle _{Z}+|5\rangle _{Y}|1\rangle _{Z}+|6\rangle
_{Y}|16\rangle _{Z} 
\]

\begin{equation}
+|7\rangle _{Y}|36\rangle _{Z}+|8\rangle _{Y}|26\rangle _{Z}+|9\rangle
_{Y}|31\rangle _{Z}+|10\rangle _{Y}|1\rangle _{Z}+|11\rangle _{Y}|16\rangle
_{Z}+|12\rangle _{Y}|36\rangle _{Z}+...+|2^{y}-1\rangle
_{Y}|f_{2^{y}-1}\rangle _{Z}\}.
\end{equation}
As Eq. (38) illustrates, the sequence of register $Z$ basis wave functions $%
|1\rangle _{Z},|f_{1}\rangle _{Z},|f_{2}\rangle _{Z},...,|f_{2^{y}-1}\rangle
_{Z}$ in Eq. (37) manifests the same periodicity with $r$ as does its
originating sequence $f_{j},$ 0 $\leq j\leq $ $2^{y}-1.$

\subsubsection{Measure the State of Register $Z.$}

The entire computer now is in the state represented by $\Psi _{C}^{2S}$ of
Eq. (37). The objective of the next three steps in the algorithm is to
extract the value of $r$ from the just discussed periodicity of $\Psi
_{C}^{2S}.$ Note that although we know $\Psi _{C}^{2S}$ has the form given
in Eq. (37), until we begin making measurements on register $Z$ we can have
no idea of what values of $f_{j}$ actually are appearing in Eq. (37).
Moreover the wave function collapse discussed under Subheading III.B.2 means
that.any single measurement on register $Z$, though it surely will reveal
one of the values of $f_{j}$ appearing in Eq. (37), automatically will
destroy all information about the other values of $f_{j}$. Nevertheless the
next step in the algorithm is to measure the state of register $Z.$ Suppose
this register $Z$ measurement on the computer in the state represented by
Eq. (37) yields the particular value $f_{k}$ (of the $r$ possible values $%
f_{0}=1,,f_{1},$ $f_{2},...,$ $f_{r-1}).$ Then after the measurement the
wave function of register $Y$ takes its third stage form 
\begin{equation}
\Psi _{Y}^{3S}=Q^{-1/2}\sum_{b=0}^{Q-1}|k+br\rangle _{Y},
\end{equation}
where, again as discussed under Subheading III.B.2: Eq. (39) has retained
those and only those $|j\rangle _{Y}$ in Eq. (37) which are multiplied by $%
|f_{k}\rangle _{Z};$ $Q$ equals the number of terms in Eq. (37) containing
the factor $|f_{k}\rangle _{Z};$ and the factor $Q^{-1/2}$ is necessary to
guarantee the wave function $\Psi _{Y}^{3S}$ of Eq. (39) is normalized,
consistent with the Born rule.$^{1}$

To help comprehend the structure of Eq. (39) and to see how the magnitude of 
$Q$ is estimated, let us return to our $N$ = 55, $n$ =16, $r=5$ illustrative
case. Suppose the result of the register $Z$ measurement on the computer in
the state represented by Eq. (38) had been $f_{j}=$ 36. Then after the
measurement the wave function of register $Y$ in this third stage of the
operation of the algorithm, was 
\begin{equation}
\Psi _{Y}^{3S}=Q^{-1/2}\{|2\rangle _{Y}+|7\rangle _{Y}+|12\rangle
_{Y}+...+|2+5(Q-1)\rangle \}.
\end{equation}
Evidently the measurement has shifted the dependence on $r,$ from the
periodicity with $r$ of the seqence $|f_{j}\rangle _{Z}$ in Eq. (37), to the
periodicity of an arithmetic progression (with common difference $r)$ of the
integers $j=k+br$ indexing the computational basis wave functions $|j\rangle
_{Y}$ appearing in Eq. (39). It is evident from Eq. (40) that the value of Q
in Eq. (39) is determined by the condition that $k+r(Q-1)$ cannot exceed 2$%
^{y}-1,$ the largest $j$ appearing in Eq. (37). Since 0 $\leq k$ 
\mbox{$<$}%
$r,$ and Q is an integer by definition, this condition implies 
\begin{equation}
Q=[r^{-1}(2^{y}-1-k)]+1,
\end{equation}
with [$x]$ denoting the largest integer $\leq x,$ as previously. We see that
unless 2$^{y}/r$ is an integer, $Q$ in Eq. (39) equals either $[2^{y}/r]$ or 
$[2^{y}/r]+1,$ depending on the value of $k;$ for the large $N$ cases of
interest herein, either of these two possible values of $Q$ is very well
approximated by 2$^{y}/r,$ because $r<N/2$ (see Subsection IV.B) whereas 2$%
^{y}>$ $N^{2}$ (as noted under Subheading III.C.1). When 2$^{y}/r$ is an
integer, however, i.e., when $r$ happens to be a power of 2 (as can happen,
recall the illustrative $f_{j}$ sequence $N$ = 55, $n=12$ discussed in
Subsection III.A), Eq. (41) makes $Q$ exactly equal to 2$^{y}/r$ (no
approximation needed) for every allowed value of $k.$

\subsubsection{Perform a Quantum Fourier Transform Operation on the Register 
$Y$ Wave Function.}

The desired $r$ finally is extracted from $\Psi _{Y}^{3S}$ of Eq. (39) via a 
$quantum$ $Fourier$ $transform$ operation {\bf U}$_{FT}.$ The operation {\bf %
U}$_{FT}$ transforms any state $|j\rangle _{Y}$ in register $Y$ to 
\begin{equation}
{\bf U}_{FT}|j\rangle _{Y}=\left( \frac{1}{\sqrt{2}}\right)
^{y}\sum_{c=0}^{2^{y}-1}e^{2\pi ijc/2^{y}}|c\rangle _{Y}.
\end{equation}
After the operation {\bf U}$_{FT}$, therefore, the wave function for
register $Y$ takes its fourth stage form 
\begin{equation}
\Psi _{Y}^{4S}={\bf U}_{FT}\Psi
_{Y}^{3S}=(2^{y}Q)^{-1/2}\sum_{c=0}^{2^{y}-1}\sum_{b=0}^{Q-1}e^{2\pi
i(k+br)c/2^{y}}|c\rangle _{Y}.
\end{equation}
It has been shown$^{54,59}$ that {\bf U}$_{FT}$ can be written as a product
of universal quantum gates, implying that {\bf U}$_{FT}$ is unitary, as we
know it is required to be. The number of gates required is$^{54,59}$ $%
O(L^{2}).$

In Eq. (43) the coefficient of any given $|c\rangle _{Y}$ is a geometric
series, i.e., is trivially summable. Performing the sum yields 
\begin{equation}
\Psi _{Y}^{4S}=(2^{y}Q)^{-1/2}\sum_{c=0}^{2^{y}-1}e^{2\pi ikc/2^{y}}\frac{
1-e^{2\pi ircQ/2^{y}}}{1-e^{2\pi irc/2^{y}}}|c\rangle
_{Y}=(2^{y}Q)^{-1/2}\sum_{c=0}^{2^{y}-1}e^{2\pi ikc/2^{y}}e^{\pi
irc(Q-1)/2^{y}}\frac{\sin (\pi rcQ/2^{y})}{\sin (\pi rc/2^{y})}|c\rangle
_{Y}.
\end{equation}

\subsubsection{Measure the State of the $Y$ Register.}

This measurement will find the $Y$ register in some particular state $%
|c\rangle _{Y}.$ The probability $P_{c}$ of finding the state $|c\rangle
_{Y} $ is given by the square of the absolute value of the coefficient of $%
|c\rangle _{Y}$ in Eq. (44), namely 
\begin{equation}
P_{c}=(2^{y}Q)^{-1}\frac{\sin ^{2}(\pi rcQ/2^{y})}{\sin ^{2}(\pi rc/2^{y})}.
\end{equation}
It is worth remarking that because the step 5 operation {\bf U}$_{FT}$ does
not involve the $Z$ register, the same Eq. (45) for the probability of
finding the $Y$ register in the state $|c\rangle _{Y}$ would obtain even if
the step 4 measurement of the $Z$ register's state had been postponed to the
present step, i.e., even if the states of both registers had been
simultaneously measured after performance of the quantum Fourier transform
operation, as Shor$^{3}$ and Volovich$^{54}$ prescribe. For this paper's
pedagogical purpose, however, it is preferable to measure the states of the
two registers in two separate steps, as Williams and Clearwater$^{56}$ also
recognize.

In order to grasp the implications of Eq. (45), it is helpful to consider
first the exceptional case that the order $r$ is a power of 2. In this
circumstance, $Q$ exactly equals $2^{y}/r$ as explained under Subheading
III.C.4. Correspondingly Eq. (45) becomes 
\begin{equation}
P_{c}=(2^{y}Q)^{-1}\frac{\sin ^{2}\pi c}{\sin ^{2}(\pi rc/2^{y})}.
\end{equation}
Because $c$ is an integer 0 $\leq $ $c$ $\leq $ 2$^{y}-1,$ Eq. (46) implies $%
P_{c}=0$ for any $c$ other than values of $c$ for which $rc/2^{y}$ is an
integer $d$, as can occur since 2$^{y}/r$ now is an integer. For such
exceptional values of $c,$ namely the values of $c$ for which 
\begin{equation}
\frac{c}{2^{y}}-\frac{d}{r}=0,
\end{equation}
the right side of Eq. (46) becomes 0/0 and we have to return to Eq. (43),
wherein we see that except for the common factor $e^{2\pi ikc/2^{y}}$ every
term in the sum over $b$ for given $c$ is unity. The number of terms in the
sum is $Q.$ So when $r$ is a power of 2 and the $Y$ register is in the state
described by the wave function $\Psi _{Y}^{4S}$ of Eq. (43), the probability 
$P_{c}$ that the $Y$ register will be found in the basis state $|c\rangle
_{Y}$ is zero except when $c$ satisfies Eq. (47), in which event $%
P_{c}=(2^{y}Q)^{-1}Q^{2}=1/r.$ Moreover, since $c$ 
\mbox{$<$}%
2$^{y},$ the only values of $c$ that can be observed are those corresponding
to the integers $d$ in the range 0 $\leq d<r;$ thus the total probability of
observing these values of $c$ is $rP_{c}=1,$ as of course it must be.

Consider now the more general circumstance that the order $r$ is not purely
a power of 2. Then there no longer exist values of $c$ which satisfy Eq.
(47) for every integer $d,$ 0 $\leq $ $d$ $<r;$ in fact if $r$ is odd one
sees there are no values of $c$ satisfying Eq. (47). Also we know from the
discussion under Subheading III.C.4 that $Q$ - 2$^{y}/r$ now equals a
non-integer $\xi ,$ where -1 
\mbox{$<$}%
$\xi $ 
\mbox{$<$}%
1. Accordingly, when $r$ is not a power of 2 the numerator in Eq. (45) is
not zero except possibly at a limited number of very special values of $c;$
in other words for most, perhaps all values of c, $P_{c}$ now is not zero$.$
Nevertheless for each integer $d$ in the allowed range the probability of
observing the result $c$ in a measurement on the $Y$ register remains large
for, and only for, those exceptional values of $c$ which--though no longer
satisfying Eq. (47)--come close to doing so. To quantify this assertion note
first that because $r$ 
\mbox{$<$}%
$N/2$, the maximum allowed value of $d/r$ (namely 1 - 1/$r)$ surely is less
than the maximum allowed value of $c/2^{y}$ (namely 1 - 1/2$^{y}$ which is $%
> $ 1 - 1/$N^{2}).$ Thus, since the spacing between successive values of $%
c/2^{y}$ is 2$^{-y},$ now every allowed value of $d/r$ either exactly
satisfies Eq. (47) for some value of $c$ or else differs from some $c/2^{y}$
by no more than 2$^{-y}/2.$ In other words when $r$ is not purely a power of
2 Eq. (47) is replaced by 
\begin{equation}
\left| \frac{c}{2^{y}}-\frac{d}{r}\right| \leq 2^{-y}/2,
\end{equation}
with the assurance that for each allowed value of $d/r$ there exists a
single c = $c_{1}$ satisfying Eq. (48) (except when the equality holds for
such a $c_{1}$, in which event the equality holds for a second $c=c_{2}$ $=$ 
$c_{1}\pm 1,$ corresponding to $d/r$ lying exactly halfway between two
successive values of $c/2^{y}).$ When $c$ satisfies Eq. (48), therefore, we
have 
\begin{equation}
\frac{c}{2^{y}}=\frac{d}{r}+\varepsilon 2^{-y},
\end{equation}
where -1/2 $\leq $ $\varepsilon $ $\leq $ 1/2.

Employing Eq. (49) in Eq. (45), the probability of finding the $Y$ register
in this $|c\rangle _{Y}$ state now (when $r$ no longer is a power of 2) is
seen to be 
\begin{equation}
P_{c}=(2^{y}Q)^{-1}\frac{\sin ^{2}(\pi r\varepsilon Q/2^{y})}{\sin ^{2}(\pi
r\varepsilon /2^{y})}\geq \frac{Q}{2^{y}}\frac{\sin ^{2}(\pi r\varepsilon
Q/2^{y})}{(\pi r\varepsilon Q/2^{y})^{2}},
\end{equation}
using the fact that $\sin x\leq x;$ the equality in Eq. (50) holds only when 
$\varepsilon =0.$ Because $2^{y}$ is so very large compared to both unity
and $r$ 
\mbox{$<$}%
$N/2,$ estimating the right side of Eq. (50) via replacement of $Q$ by $%
2^{y}/r$ (although $Q$ actually differs from $2^{y}/r$ by a quantity $\xi ,$ 
$\left| \xi \right| <1)$ can be seen to introduce an inconsequential error;
in other words the argument of the sine function on the right side of Eq.
(50) may be taken to be $\pi \varepsilon .$ Hence Eq. (50) yields 
\begin{equation}
P_{c}\geq r^{-1}\frac{\sin ^{2}\pi \varepsilon }{(\pi \varepsilon )^{2}}\geq
r^{-1}\frac{1}{(\pi /2)^{2}}=\frac{4}{r\pi ^{2}},
\end{equation}
where the second inequality results from recognizing that $\sin x/x$ is a
decreasing function of $x$ in the range 0$\leq x\leq \pi $, and then
replacing $\left| \varepsilon \right| $ by its maximum allowed value 1/2.
Since there is such a $c$ and associated $P_{c}$ for each of the $r$ allowed
values of $d$ in Eq. (48), we conclude that even when $r$ is not a power of
2 the total probability $P$ = $rP_{c}$ of finding the $Y$ register in a
state $|c\rangle _{Y}$ for which $c$ satisfies Eq. (48) is not less than 4/$%
\pi ^{2}$ $\cong $ 0.4.

This just stated result for $P$ has been obtained by Ekert and Josza$^{4}$;
it is larger than the value of $P$ originally quoted by Shor.$^{3}$ It is
clear from its derivation, however, that this lower bound of 0.4 (though
rigorously derived) considerably underestimates the magnitude of $P$ that is
likely to be encountered in practice. For example, if in Eq. (51) $\left|
\varepsilon \right| $ is replaced not by its maximum value but rather by its
average value 1/4, then Eq. (51) yields $P_{c}\geq $ 8/$r\pi ^{2},$
corresponding to $P\geq 0.81$. Use of the average value of $\left|
\varepsilon \right| $ to estimate $P$ is reasonable because in general the
value of $\varepsilon $ depends on $d,$ as can be seen from Eq. (49)
recollecting that 2$^{y}/r$ is not an integer unless $r$ is a power of 2.

\subsubsection{Infer the Value of $r.$}

After a value of $c$ has been obtained, i.e., after the state measurement on
register $Y$ prescribed in the previous step, it still is necessary to infer
the value of $r.$ To help understand how this inference is accomplished, I
observe first that, because $r$ is less than $N/2,$ there can be only one
permissible fraction $d/r$ satisfying Eq. (48) for any given $c;$ here
''permissible'' means of course that $d$ is an integer 0 
\mbox{$<$}%
$d<r<N/2.$ To prove this assertion note that if $d_{1}/r_{1}$ and $%
d_{2}/r_{2}$ are two distinct permissible fractions, i.e., if $%
d_{1}/r_{1}\neq d_{2}/r_{2},$ then

\begin{equation}
\left| \frac{d_{1}}{r_{1}}-\frac{d_{2}}{r_{2}}\right| =\left| \frac{
d_{1}r_{2}-d_{2}r_{1}}{r_{1}r_{2}}\right| >\frac{1}{r_{1}r_{2}}>\frac{4}{%
N^{2}},
\end{equation}
since when $d_{1}/r_{1}$ $\neq $ $d_{2}/r_{2},$the integer ($%
d_{1}r_{2}-d_{2}r_{1})$ cannot equal 0. On the othe hand if these $%
d_{1}/r_{1}$ and $d_{2}/r_{2}$ each satisfy Eq. (48) for the same $c,$

\begin{equation}
.\left| \frac{d_{1}}{r_{1}}-\frac{d_{2}}{r_{2}}\right| =\left| (\frac{c}{%
2^{y}}-\frac{d_{2}}{r_{2}})-(\frac{c}{2^{y}}-\frac{d_{1}}{r_{1}})\right|
\leq \left| \frac{c}{2^{y}}-\frac{d_{2}}{r_{2}}\right| +\left| \frac{c}{2^{y}%
}-\frac{d_{1}}{r_{1}}\right| \leq 2(2^{-y}/2)=2^{-y}<\frac{1}{N^{2}}.
\end{equation}
Since Eqs. (52) and (53) are inconsistent, the impossibility of finding two
distinct $d/r$ satisfying Eq. (48) for the same $c$ is proved..

Suppose now that our state measurement on the $Y$ register has yielded a $%
|c\rangle _{Y}$ state whose $c$ satisfies Eq. (48). The actual evaluation of
this (now known to be unique) $d/r,$ and thence $r,$ from Eq. (48) is
performed by expanding $c/2^{y}$ into a continued fraction, as Shor$^{3}$
originally proposed. I shall not detail herein the construction of continued
fraction expansions; such quite comprehensible discussions are readily found$%
^{60},$ and I do provide an illustrative expansion below as well as (under
Subheading IV.D.2) an explanation of the relation between continued fraction
expansions and gcd calculations. Suffice it to say that the continued
fraction expansion of any rational number $x$ provides a series of fractions
(with each fraction in lowest terms) called $convergents$ to $x,$ such that
each successive convergent furnishes an improved approximation to $x.$ A key
theorem$^{61}$ is: Let $a/b$ be a fraction satisfying 
\begin{equation}
\left| \frac{a}{b}-x\right| <\frac{1}{2b^{2}}.
\end{equation}
Then $a/b$ is one of the continued fraction convergents to $x.$ Eq. (48) has
the form of Eq. (54), with $x$ $\equiv $ $c/2^{y}$ and $a/b$ $\equiv $ $d/r;$
since 2$^{y}>N^{2},$ the right side of Eq. (48) is less than (2$N^{2})^{-1},$
which in turn is less than (2$r^{2})^{-1}$ because $r$ 
\mbox{$<$}%
$N/2.$ Therefore, by this just quoted theorem, $d/r$ must be one of the
continued fraction convergents to $c/2^{y},$ i.e., expanding $c/2^{y}$ in
its series of continued fraction convergents inevitably will yield $d/r$ in
lowest terms. Note that this result demonstrates the critical importance of
choosing the size $y$ of the $Y$ register 
\mbox{$>$}%
\mbox{$>$}%
$N.$ Indeed, if 2$^{y}<N^{2}/4,$ the right side of Eq. (48) would be $>$ 2/$%
N^{2},$ i.e., no longer would ensure that $d/r$ is one of the continued
fraction convergents to $c/2^{y}.$ Similarly, if 2$^{y}$ 
\mbox{$<$}%
$N^{2}/4,$ the right side of Eq. (53) would be 
\mbox{$>$}%
4/$N^{2},$ i.e., Eq. (53) no longer would be inconsistent with Eq. (52),
implying that it now no longer is guaranteed there is only one permissible $%
d/r$ satisfying Eq. (48).

Let me illustrate this beautifully simple continued fraction method of
determining $d/r.$ To factor our illustrative $N=55$ via Shor's algorithm a $%
Y$ register of size $y=12$ qubits will be employed, as prescribed under
Subheading III.C.1 (2$^{11}=2048$ 
\mbox{$<$}%
$N^{2}=3025<2^{y}=4096)$. For this $N$ the order of $n=37$ is $r=20,$ the
largest possible value of $r$ for this $N$ (as stated under the same
Subheading). For $d=11,$ the value of $d/r$ is exactly 0.55. We have
2252/4096 = 0.54980; 2253/4096 = 0.55005; and 2$^{-y}/2=0.00012,$ which is 
\mbox{$<$}%
0.00020 = 0.55 - 2252/4096, but is 
\mbox{$>$}%
0.00005 = 2253/4096 - 0.55. Then if we assume the state measurement on the $%
Y $ register has yielded the state $|c\rangle _{Y}$ consistent with Eq. (48)
for $r$ = 20 and $d=11,$ the value of $c$ must have been 2253. We now expand
2253/4096 in a continued fraction: 
\begin{equation}
\frac{2253}{4096}=\frac{1}{\frac{4096}{2253}}=\frac{1}{1+\frac{1843}{2253}}=%
\frac{1}{1+\frac{1}{\frac{2253}{1843}}}=\frac{1}{1+\frac{1}{1+\frac{410}{1843%
}}}=\frac{1}{1+\frac{1}{1+\frac{1}{\frac{1843}{410}}}}=\frac{1}{1+\frac{1}{1+%
\frac{1}{4+\frac{203}{410}}}}=\frac{1}{1+\frac{1}{1+\frac{1}{4+\frac{1}{2+%
\frac{6}{203}}}}}.
\end{equation}
and so on. Dropping the fraction 410/1843 in the expression to the right of
the fourth equality sign in Eq. (55) yields the first convergent, namely
1/2; dropping the fraction 203/410 in the expression to the left of the last
equality in Eq. (55) yields the second convergent, namely 5/9 = 0.5555. Each
of these two convergents differs from 2253/4096 by an amount whose absolute
value exceeds 0.00012, i.e., each of these convergents fails to satisfy Eq.\
(48) and so cannot equal the desired $d/r.$ Lo and behold, however, the
third convergent, obtained by dropping the fraction 6/203 in the expression
to the right of the last equality in Eq. (55), is precisely 11/20,
confirming the theorem quoted in the preceding paragraph. Moreover, because
we know $r$ 
\mbox{$<$}%
$N/2,$ which in this illustrative case is 55/2, the result that $d/r$ =
11/20 immediately implies that $r$ = 20, because any other fraction equal to
11/20, e.g., 22/40, inevitably has a denominator 
\mbox{$>$}%
55/2.

I next observe that because $r<N/2,$ not merely 
\mbox{$<$}%
$N,$ it follows from Eq. (54) that even if the right side of Eq. (48) had
been replaced by 2/$N^{2},$ values of $c/2^{y}$ satisfying the thus modified
Eq. (48) would have continued fraction convergents equal to $d/r.$ But 2$%
^{y} $ 
\mbox{$>$}%
$N^{2}$ implies 2/2$^{y}<2/N^{2}.$ In other words if a state measurement on
the $Y$ register should yield a $|c\rangle _{Y}$ whose $c$ satisfies 
\begin{equation}
\left| \frac{c}{2^{y}}-\frac{d}{r}\right| \leq 2(2^{-y}),
\end{equation}
that $c/2^{y}$ also will have $d/r$ as a continued fraction convergent of $%
c/2^{y}$ even though the value of $c$ may not satisfy Eq. (48)$.$ Therefore
we now have another reason (in addition to the previously explained
desirability of using an average $\left| \varepsilon \right| )$ for
asserting that the quantity 4/$\pi ^{2}$ quoted following Eq. (51) greatly
underestimates the probability of measuring states $|c\rangle _{Y}$ which
can yield $d/r.$ To more accurately estimate this probability, note that if $%
c/2^{y}>d/r$ satisfies Eq. (48), then each of $c-2,$ $c-1$, $c$ and $c+1$
will satisfy Eq. (56). Similarly, if $c/2^{y}<d/r$ satisfies Eq. (48), then
each of $c-1,$ $c,$ $c+1$ and $c+2$ will satisfy Eq. (56). In either case,
adding the appropriate four $P_{c}$ from Eq. (45), using $\sin x\leq x$ as
in Eq. (50), and approximating $Q$ by 2$^{y}/r$ as was done in deriving Eq.
(51), we find that the probability $P_{c}^{\prime }$ of measuring a state $%
|c\rangle _{Y}$ which will have $d/r$ as a continued fraction convergent of $%
c/2^{y}$ is 
\begin{equation}
P_{c}^{\prime }\geq \frac{\sin ^{2}\pi \varepsilon }{\pi ^{2}r}\left( \frac{1%
}{(1+\varepsilon )^{2}}+\frac{1}{\varepsilon ^{2}}+\frac{1}{(1-\varepsilon
)^{2}}+\frac{1}{(2-\varepsilon )^{2}}\right) ,
\end{equation}
where now 0 $\leq $ $\varepsilon \leq 1/2,$ and the prime on $P_{c}^{\prime
} $ indicates that we have summed over the appropriate four $P_{c}$ as
explained above. For $\varepsilon $ = 1/2, we obtain $P_{c}^{\prime }=$ 80/9$%
\pi ^{2}r=0.90/r;$ using the average $\varepsilon $ = 1/4 we obtain $%
P_{c}^{\prime }=0.935/r.$ Returning to our illustrative continued fraction
expansion, it is readily verified that each of the continued fraction
expansions of 2251/4096, 2252/4096 and 2254/4096, like the Eq. (55)
expansion of 2253/4096, do have 11/20 as a convergent, consistent with our
employment of Eq. (57) to estimate the probability of correctly inferring $%
d/r$ from a state measurement $|c\rangle _{Y}.$

\subsubsection{Repeat Steps 2 Through 7 Until Factorization of $N$ is
Achieved.}

Inferring the value of $r$ need not immediately lead to factorization of $N,$
however. In the first place, as was mentioned in Subsection III.A, the
probability that $r$ will meet the necessary requirements for being able to
factor $N,$ namely that $r$ is even and satisfies Eq. (14), is$^{44}$ only
about 1/2. Thus although the probability of being able to infer a $d/r$ via
a single measurement of the $Y$ register is so high, namely over 90\%,
nevertheless on the average it will be necessary to run through the entire
sequence of steps 2 through 7 at least twice before a $d/r$ whose $r$ can be
employed to factor $N$ is obtained. The entire sequence must be repeated
starting from step 2 (we don't have to make any new decisions about the
sizes of the registers) because after step 7 the $Y$ register is in whatever
state $|c\rangle _{Y}$ was measured. The wave function of this state is
nothing like the initial loading wave function $\Psi _{Y}^{1S}$ of Eq. (31)
from which the Shor algorithm operations departed, beginning with step 3;
also, unless we already have cleared register $Z$ to the state $|0\rangle
_{Z},$ the operations described in steps 2 and 3 will not yield the desired $%
\Psi _{C}^{2S}$ of Eq. (37). In these repetitions, although the $Y$ register
wave function always will be brought to its initial loading form Eq. (31),
i.e., although step 2 always will be the same, the choice of $n$ in step 3
had better be different; otherwise carrying the algorithm through to step 7
merely will again yield an $r$ which cannot be employed to factor $N.$

Furthermore, even granting that the $n$ selected in step 3 does possess an
order $r$ which is employable to factor $N,$ inferring $r$ from the computed 
$d/r$ may not be as simple as the discussion under the immediately preceding
Subheading has suggested. Suppose, again returning to our illustrative $N$ =
55, $n$ = 37, $r$ = 20 example, the register $Y$ state measurement had
yielded $c$ = 2048, which for $r$ = 20 satisfies Eq. (48) with $d$ = 10. But
the computer operator doesn't know $r$ = 20; all the operator knows is that
2048/4096 = 1/2, the sole convergent (which has to be in lowest terms
remember) to 2048/4096. The operator immediately will discover (37)$^{2}$ $%
\equiv $ 49 $\neq $ 1 (mod 55), so that 2 surely is not the order of 37, but
then what? Each of the fractions 2/4, 3/6, 4/8,..., 13/26, equals 1/2 and
has a denominator 
\mbox{$<$}%
55/2, i.e., each of these denominators could be the desired $r.$ In
principle the operator could test the powers (37)$^{2},$ (37)$^{4},$ (37)$%
^{6},...$ (mod 55) until he/she came to (37)$^{20}$ $\equiv $ $1$ (mod 55).
For the large $N$ of interest, however, e.g., RSA-309, persistently trying
to determine $r$ in this crude fashion after the register $Y$ measurement
has yielded a convergent with a denominator $b$ for which $b$ 
\mbox{$<$}%
\mbox{$<$}%
$N$ and $n^{b}\not{\equiv}1$ $(%
\mathop{\rm mod}%
$ $N),$ obviously would be ridiculous and would negate the whole point of
using Shor's algorithm. Shor$^{3}$ suggests the operator should try a few
small multiples of $b,$ e.g., 2$b$ and 3$b;$ but after finding $n^{2b}$ and $%
n^{3b}\not{\equiv}1$ $(%
\mathop{\rm mod}%
$ $N)$, the operator seemingly would have little choice but to repeat steps
2 through 7 (this time using the same value of $n$ of course), in the hope
that the newly measured $c/2^{y}$ would have a convergent whose denominator
actually was $r,$ not a factor of $r.$

How many times the operator may expect to have to repeat steps 2 through 7
before reliably inferring $r$ (still assuming the operator has selected an $%
n $ possessing an employable $r)$ is difficult to say. A seeming
overestimate of the expected number of such repetitions follows from
considerations first advanced by Shor$^{3}$ and refined by Ekart and Josza.$%
^{4}$ The number of positive integers $d$ less than $r$ that are coprime to $%
r$ is $\phi (r),$ where $\phi $ is Euler's $totient$ $function^{62}$ (the
subject of Subsection IV.B). Then if $P^{\prime }$ is the probability (equal
to at least 0.9 as we have seen) that a measurement on the $Y$ register will
yield a $c/2^{y}$ with a convergent equal to some $d/r$, 0 $\leq $ $d$ 
\mbox{$<$}%
$r,$ then $P^{\prime \prime }=$ $P^{\prime }\phi (r)/r$ is the probability
that the measurement will yield a convergent equal to a $d/r$ wherein $d$ is
prime to $r.$ Ekart and Josza$^{4}$ quote the theorem$^{63}$ that for
sufficiently large $r$ 
\begin{equation}
\frac{\phi (r)}{r}\geq \frac{0.56}{\ln \ln r}\cong \frac{1.17}{\log _{2}\log
_{2}r}.
\end{equation}
Because the typical $r$ is expected to increase as $N$ increases, Eq. (58)
suggests that whatever may be the number of repetitions 2 through 7
otherwise required (e.g., repetitions because $r$ is not always employable
to factor $N)$ those repetitions might need to be increased by a factor of
about $\log _{2}\log _{2}r$ because of the just discussed complications
associated with fractions $d/r$ in Eq. (56) wherein $d$ is not coprime to $%
r. $

This just estimated increase in the required number of repetitions probably
is an overestimate because it does not take into account the likelihood (as
explained in the penultimate paragraph) that the operator will infer $r$
without recourse to repetitions when $r$ is only a small multiple of the
denominator $b$ of the measured convergent$,$ e.g., when $b$ equals $r/2$ or 
$r/3.$ The operator also can minimize the number of required repetitions by
recognizing (as Shor$^{3}$ also remarks) that if two measured convergents
have denominators $b_{1}$ 
\mbox{$<$}%
\mbox{$<$}%
$N/2$ and $b_{2}$ 
\mbox{$<$}%
\mbox{$<$}%
$N/2$ with $b_{1}$ coprime to $b_{2},$ then the only way for $r$ to be a
multiple of each of $b_{1}$ and $b_{2}$ is for $r$ to be a multiple of $%
b_{1}b_{2},$ which now may be sufficiently large to ensure that $r$ is
either 2$b_{1}b_{2.}$ or 3$b_{1}b_{2.}.$ For instance, returning once again
to our illustrative $N$ = 55, $n$ = 37, $r$ = 20 example, if after obtaining
the convergent 1/2 the operator were to repeat steps 2 through 7 with the
same $n,$ and if this repeat should yield the convergent 3/5, the operator
would infer with high probability (greater than 0.9 as discussed above) that 
$r$ is a multiple of 10; once having discovered that (37)$^{10}$ $\equiv $ $%
34$ (mod 55), the operator would infer with the same high probability that $%
r $ = 20, because 30 = 3(10) is 
\mbox{$>$}%
55/2 and therefore cannot be $r.$ Indeed once having verified that (37)$%
^{20} $ $\equiv $ $1$ (mod 55), the knowledge that (37)$^{10}$ $\equiv $ $34$
(mod 55) immediately enables the factors 5 and 11 of $N=55$ to be
determined, precisely as was illustrated at the end of Subsection III.A. \ \
\ 

\subsection{Concluding Remarks.}

The foregoing completes this paper's presentation of the operations
constituting Shor's algorithm. The following added remarks seem appropriate,
however. The algorithm involves the application of unitary operations at
steps 2, 3 and 5. The estimated numbers of gates required to accomplish each
of these steps are stated in the text under their respective subheadings;
denote these estimated numbers by $\nu _{q2},$ $\nu _{q3}$ and $\nu _{q5}$
respectively. The estimated total number of gates required, denoted by $\nu
_{q}$ in Eq. (28), equals $\nu _{q2}+\nu _{q3}+\nu _{q5}$. We see that in
the limit of very large $N$ the estimates $\nu _{q2}$ and $\nu _{q5}$ become
negligible compared to $\nu _{q3}.$ Accordingly Eq. (28) equates $\nu _{q}$
to $\nu _{q3},$ in agreement with conventional procedure.$^{3,54}$
Admittedly Eq. (28) has not taken into account the operations, gate or
otherwise, required to perform the state measurements postulated under the
algorithm steps 4 and 6. We have seen that the quantum computer can carry
out the algorithm with no more than about 3L qubits, however; it is
difficult to see why the required number $\nu _{m}$ of measurement
operations, including the post-measurement operations needed to restore the
computer wave function to its (for the purpose of the algorithm, recall the
discussion under Subheading III.C.2) starting form $\Psi _{C}^{(0)}$ = $%
|0\rangle _{Y}|0\rangle _{Z}$ should be other than proportional to the
number of qubits. Consequently the failure to include state measurement
operations in no way invalidates employing Eq. (28), which grows somewhat
faster than $L^{2}$ with increasing $N,$ to estimate the growth with $N$ of
the computing effort required to perform a factorization of $N$ using Shor's
algorithm. $.$

If in practice repetitions of the algorithm steps are necessary, as the
discussion under the step 8 Subheading indicates almost always will be the
case, then those repetitions (each of which necessitates a new setting up of
the gates) should be taken into account in any estimate, such as Eq. (28),
of the total number of gates required to determine an order $r$ permitting
factorization of $N.$ Because the $Order$ $Of$ symbol $O$ defined in Eq.
(29) has been included in Eq. (28), any required number of repetitions which
does not increase with $N$ (e.g., the expected number of repetitions
associated with the fact that some $r$ will not be employable to factor $N),$
does not demand any correction of Eq. (28). On the other hand the number of
repetitions suggested by Eq. (58), although very likely a considerable
overestimate of the actual number of required repetitions associated with
the desirability of measuring a $d/r$ for which $d$ is prime to $r,$
probably does demand some modification of Eq. (28). If we assume that the
typical $r<N/2$ tends to be some fixed fraction of $N,$ then for large $N$
we can replace $\log _{2}\log _{2}r$ with $\log _{2}\log _{2}N,$ therewith
concluding that our earlier text concerning Eq. (28) should have been
supplemented by: An upper bound $\nu _{qub}(N)$, on the expected number of
universal quantum gates that actually will have to be employed in a Shor
algorithm determination of an order $r$ permitting factorization of $N$ is
obtained via multiplication of $\nu _{q}(N)$ in Eq. (28) by log$_{2}L,$
yielding

\begin{equation}
\nu _{qub}(N)=O[L^{2}(\log _{2}L)^{2}(\log _{2}\log _{2}L)].
\end{equation}
Eq. (59) only very minimally weakens the conclusions drawn earlier from
comparison of Eqs. (7) and (28), or from computing the actual magnitude of $%
\nu _{q}(N)$ given by Eq. (29). For instance, whereas previously we
concluded that a quantum computer which could factor RSA-309 in two weeks
time should be able to factor RSA-617 in no more than about 9 weeks, Eq.
(59) leads to the conclusion that a quantum computer which surely can factor
RSA-309 in no more than 2 weeks will factor RSA-617 in at most 10 weeks.

After a $c$ has been measured, as described under the step 6 Subheading, the
following calculations still must be performed: (i) infer an $r$ from the
measured $c,$ which generally involves a continued fraction expansion as
discussed under the Step 7 Subheading; (ii) verify that the inferred $r$
satisfies Eqs. (10), (12) and (14) (as it must if this $r$ is to be
employable to factor $N),$ which involves computing $n^{r}$ and $n^{r/2}$
(mod $N);$ and (iii) then actually obtain the factors $p$ and $q$ of $N,$
which process involves computing greatest common divisors as discussed
immediately following Eq. (14). At present it is not contemplated that any
of these required calculations will be accomplished by any computer other
than a purely classical one. The efforts required to accomplish these
computations have not been included in Eq. (28), nor could they be, because
Eq. (28) estimates the number of universal quantum gates required, not the
number of classical computer bit operations as in Eq. (7). On the other hand
the efforts to perform these classical calculations are not irrelevant to
any realistic estimate of the potential utility of Shor's algorithm for
factoring increasingly large $N.$ It is pertinent to remark, therefore, that
(see Section IV.D) for none of the computations listed under (i) - (iii)
immediately above does the number of required bit operations increase with $%
N $ more rapidly than the right side of Eq. (59). Consequently Eq. (59)
correctly exhibits the maximum expected growth with increasing $N=pq$ of the
total computational efforts, quantum plus classical, required to complete a
factorization of $N$ using Shor's algorithm. Correspondingly, the
conclusions we have drawn from comparisons of Eqs. (7) and (28) remain
valid, except for the very minimal weakening discussed immediately beneath
Eq. (59), even though the derivation of Eq. (28) ignored the classical
computer calculations presently inherent in the use of Shor's algorithm to
factor $N.$

Until Shor produced his algorithm it was generally believed that the
computational effort required to factor $N$ = $pq$ grows more rapidly with
increasing $N$ than any polynomial in $L=\log _{2}N,$ as Eq. (7) manifests.
Shor's demonstration that use of a quantum computer could decrease this
growth to slower than $L^{3}$ was astonishing, therefore, and has greatly
accelerated attempts to construct an actually functioning quantum computer.
The key Shor algorithm operation, the operation that enables the greatly
diminished growth of the computational effort with $N,$ is the quantum
Fourier transform {\bf U}$_{FT}$ operation discussed under the step 5
Subheading. The quantum Fourier transform is a direct generalization (to
quantum mechanical basis states) of the classical computing $discrete$ $%
Fourier$ $transform,$ which in turn is nothing more than a discretized
(summation rather than integration) version of the conventional Fourier
integral transform. Thus it does not seem surprising that application of 
{\bf U}$_{FT}$ to the wave function $\Psi _{Y}^{3S}$ of Eq. (39) yields a
new wave function, namely $\Psi _{Y}^{4S}$ of Eq. (44), wherein the
probability $P_{c}$ [that a measurement on the $Y$ register will yield the
basis state $|c\rangle _{Y}]$ is large only for those values of $c$ from
which the periodicity with $r$ inherent in Eq. (40) can be inferred. What is
very remarkable, however, and what makes possible the comparatively slow
increase with $N$ of $\nu _{q}(N)$ in Eq. (28), is the fact that although
the discrete Fourier transform calculation requires $O(NL)$ bit operations$%
^{64},$ performance of {\bf U}$_{FT}$ can be accomplished with only $%
O(L^{2}) $ universal quantum gates, as stated immediately following Eq.
(43). It must be remembered that {\bf U}$_{FT}$ is a 2$^{y}\times 2^{y}$
matrix, i.e., at least an $N^{2}\times N^{2}$ unitary matrix; because an
arbitrary unitary matrix of this dimensionality contains $N^{4}$ free
parameters, in general one expects that reproducing a given $N^{2}\times
N^{2}$ unitary matrix will require a sequence of no fewer than $N^{4}/16$
one-qubit and two-qubit gates. This observation, based on trivial
dimensional considerations, suggests that for most classical computer
algorithms the growth of computational effort with number size will not be
importantly diminished merely by recasting the algorithm into a form usable
in a quantum computer.

Finally we remark that factorization of a number $N$ = $pq$ via a quantum
computer using Shor's algorithm actually has been accomplished$^{65};$
although the number factored, namely 15, is the smallest possible product of
odd primes, the accomplishment assuredly is notable. It also is notable,
however, that because $\phi (15)=4\times 2=8$ the only possible orders $r$
were $r=2$ and $r=4,$ meaning that in this quantum computer factoring
demonstration the value of $r$ could be inferred from Eq. (47) for any
chosen $n$ coprime to and 
\mbox{$<$}%
15, without the complications attendant upon the much more usual
circumstance that $r$ has to be inferred from Eq. (48). Nor could this
experiment test the feasibility of determining $r$ in the likely event that
repetitions of steps 2 through 7 of the algorithm will be required, as
discussed under Subheading III.C.8.

\section{Appendix. Pertinent Number Theoretic Results.}

This Appendix presents the various number theoretic derivations and other
results referred to in the preceding Sections of this paper. I take this
opportunity to thank Paul Reilly for numerous enlightening discussions,
especially on the RSA system. I am indebted to Joseph Burdis for carefully
checking the manuscript, including its references. I also am indebted to Sam
Scheinman for data on RSA enciphering and deciphering.

\subsection{\protect\smallskip Congruence Manipulations. Illustrative RSA
Operations. Periodicity of Remainders $f_{j}.$}

Comparison of Eqs. (1) and (5) illustrates the proposition that (subject to
the important proviso that all the congruences must have the same modulus $%
m) $ congruences like Eqs. (1)-(5) have the useful property that in many
respects they can be manipulated as if they were equalities, i.e., as if the
congruence symbol $\equiv $ were the equality symbol =. For example Eq. (1)
and 
\begin{equation}
x\equiv y\hspace{1in}(%
\mathop{\rm mod}%
\;m)
\end{equation}
imply both 
\begin{equation}
ax\equiv by\hspace{1in}(%
\mathop{\rm mod}%
\;m).
\end{equation}
and 
\begin{equation}
bx\equiv ay\hspace{1in}(%
\mathop{\rm mod}%
\;m),
\end{equation}
Accordingly Eq. (1) implies 
\begin{equation}
a^{z}\equiv b^{z}\hspace{1in}(%
\mathop{\rm mod}%
\;m).
\end{equation}
for any positive integer $z.$ Eqs. (61)-(63) can be trivially demonstrated$%
^{24}$ remembering that Eq. (1) means $a=b+wm,$ w some positive or negative
integer. There are a few permissible manipulations of equalities that have
no congruence counterparts, but any such manipulations are not relevant to
this paper. Unless explanatory comments seem required, therefore, the
remainder of this Appendix will manipulate congruences as if they were
equalities without further ado.

The use of congruence manipulations to conveniently compute Alice's $S$ =
21, 14, 51, 1, 13, 27, 10, 1, 9, 8, 49, 51 (given at the end of Subsection
II.B) from her illustrative $C=$ 21, 9, 6, 1, 7, 3, 10, 1, 4, 2, 14, 6
(quoted in Subsection II.A) will now be exemplified. Our illustrative RSA
key number and encryption exponent, to be inserted into Eq. (2) along with
each $c$ in $C$ are $N=$ 55 and $e=23$ respectively. Consider initially
Alice's first $c$ = 21. Instead of determining the corresponding $s$ by
tediously computing $c^{e}=$ (21)$^{23}$ and then dividing by 55, Alice
proceeds as follows: 
\begin{equation}
(21)^{2}=441\equiv 1\hspace{1in}(%
\mathop{\rm mod}%
\;55),
\end{equation}
\begin{equation}
(21)^{22}\equiv (1)^{11}\equiv 1\hspace{1in}(%
\mathop{\rm mod}%
\;55),
\end{equation}
\begin{equation}
(21)^{23}\equiv (21)(21)^{22}\equiv 21(1)\equiv 21\hspace{1in}(%
\mathop{\rm mod}%
\;55).
\end{equation}
So the first $s$ in $S$ turns out to equal the first $c$ = 21. The second $s$
is obtained not quite as simply, but surely a lot more easily than having to
exactly compute (9)$^{23},$ namely: 
\begin{equation}
(9)^{2}=81\equiv 26\hspace{1in}(%
\mathop{\rm mod}%
\;55),.
\end{equation}
\begin{equation}
(9)^{4}\equiv (26)^{2}=676\equiv 16\hspace{1in}(%
\mathop{\rm mod}%
\;55),
\end{equation}
\begin{equation}
(9)^{8}\equiv (16)^{2}=256\equiv 36\equiv -19\hspace{1in}(%
\mathop{\rm mod}%
\;55),
\end{equation}
\begin{equation}
(9)^{10}\equiv (-19)(26)=-494\equiv -54\equiv 1\hspace{1in}(%
\mathop{\rm mod}%
\;55),
\end{equation}
\begin{equation}
(9)^{20}\equiv (1)^{2}\equiv 1\hspace{1in}(%
\mathop{\rm mod}%
\;55),
\end{equation}
\begin{equation}
(9)^{23}\equiv (9)^{20}(9)^{2}(9)\equiv (26)(9)=234\equiv 14\hspace{1in}(%
\mathop{\rm mod}%
\;55).
\end{equation}
Thus the second $s$ is 14. Similar congruence manipulations on $C$ readily
yield the above-quoted complete $S$, as readers of this paper now readily
can verify. Similarly it is readily verified that Bob's secret decryption
exponent $d=7$ really does decipher this $S$ into $C,$ namely that in
accordance with Eq. (3): (21)$^{7}\equiv (21)(21)^{6}\equiv 21$ (mod 55),
making use of Eq. (64) above; (14)$^{2}$ = 196 $\equiv 31,$ (14)$^{4}\equiv
961\equiv 26,$ (14)$^{7}\equiv (14)(31)(26)\equiv (14)(36)\equiv 9$ (mod
55), etc.

The permissibility of these congruence manipulations also immediately
implies the periodicity of the remainders $f_{j}$ defined by Eq. (8). Using
Eq. (10), we see that $f_{j+r}$ $\equiv $ $n^{j+r}$ $\equiv $ $%
n^{j}n^{r}\equiv n^{j}\equiv $ $f_{j}$ (mod $pq),$ implying $f_{j+r}=$ $%
f_{j},$ since by definition all the $f_{j}$ are positive numbers 
\mbox{$<$}%
$N;$ similarly $f_{j+2r}\equiv n^{j+r}n^{r}\equiv n^{j+r}\equiv f_{j}$ (mod $%
pq),$ etc. It also is readily seen that all the $f_{j},$ 1 $\leq j\leq r,$
are different. For suppose $f_{a}=$ $f_{b},$ where each of $a\neq b$ lies in
the just specified range of $j.$ Suppose further $a$ 
\mbox{$<$}%
$b.$ Then $n^{b}\equiv n^{a}$ (mod $pq),$ meaning $%
n^{b}-n^{a}=n^{a}(n^{b-a}-1)$ is divisible by $pq.$ This means $n^{b-a}-1$
must be divisible by $pq,$ because $n$ has been chosen to be coprime to $pq.$
On the other hand it is not possible to have $n^{b-a}\equiv 1$ (mod $pq)$
because by definition $r$ is the smallest value of $j$ for which $%
n^{j}\equiv $ $1$ (mod $pq).$

\subsection{Euler's Totient Function. Proof That For $N=pq$ the Order $%
r<N/2. $}

For any positive integer $m,$ Euler's totient function$^{62}$ $\phi (m)$ is
the number of positive integers less than $m$ that are coprime to $m;$ by
definition $\phi (m)$ always is 
\mbox{$<$}%
$m.$ Euler proved$^{66}$ that if $a$ is coprime to $m,$ then 
\begin{equation}
a^{\phi (m)}\equiv 1\hspace{1in}(%
\mathop{\rm mod}%
\;m).
\end{equation}
Let us calculate $\phi (N)$ for key numbers $N=pq,$ where $p$ and $q$ are
odd primes. The only numbers 
\mbox{$<$}%
$N$ that are not coprime to $N$ are multiples of $p$ and $q.$ There are $q-1$
integers $p,2p,...,(q-1)p$ less than $N;$ similarly there are $p-1$
multiples of $q$ that are 
\mbox{$<$}%
$N.$ Since none of these numbers can coincide and be less than $N,$%
\begin{equation}
\phi (N)=N-1-[(p-1)+(q-1)]=pq-p-q+1=(p-1)(q-1).
\end{equation}
Evidently the RSA $\phi $ introduced in Subsection II.A, and then employed
in Eqs. (4) and (9), is $\phi (pq);$ the explicit dependence on $N=pq$ was
dropped in those equations because no possible confusion could result
therefrom. Equally evidently, Eq. (73) immediately implies Eq. (9). I note
parenthetically that if $n$ is not coprime to $pq,$ i.e., if $n$ and $pq$
have a common factor $x>1,$ then $f_{j}$ in Eq. (8) also must be divisible
by $x,$ as is immediately seen remembering Eq. (8) means $n^{j}-f_{j}=ypq,$ $%
y$ some integer. Consequently Eq. (10) cannot hold for any integer $r$
unless $n$ actually is coprime to $pq.$

If $n$ is coprime to $pq$ moreover, Eq. (9) is supplemented by 
\begin{equation}
n^{(p-1)(q-1)/2}\equiv 1\hspace{1in}(%
\mathop{\rm mod}%
\;pq),
\end{equation}
where, as in Eq. (9) and always herein, $n$ is coprime to $pq.$ To prove Eq.
(75) it is convenient to start from the the form taken by Eq. (73) when $m$
is an odd prime $p.$ Evidently $\phi (p)=p-1,$ so that 
\begin{equation}
a^{p-1}\equiv 1\hspace{1in}(%
\mathop{\rm mod}%
\;p),
\end{equation}
where $a$ is coprime to $p,$ of course. Eq. (76) is known as Fermat's $%
Little $ $Theorem,$ stated by him$^{67}$ in 1640. Because $q$ also is an odd
prime, ($q-1)/2$ is an integer, so that Eq. (76) implies [recall Eq. (63)] 
\begin{equation}
a^{(p-1)(q-1)/2}\equiv 1\hspace{1in}(%
\mathop{\rm mod}%
\;p).
\end{equation}
But if $a$ also is coprime to $q,$ then it similarly is true that 
\begin{equation}
a^{(q-1)(p-1)/2}\equiv 1\hspace{1in}(%
\mathop{\rm mod}%
\;q).
\end{equation}
If $a$ is coprime to both $p$ and $q,$ however, then $a$ is coprime to $pq,$
i.e., $a$ is an $n$ as defined at the outset of Subsection III.A. For any
positive integer $z,$ furthermore, if $z-1$ is separately divisible by a
prime $p$ and by another prime $q,$ then $z-1$ is divisible by the product $%
pq.$ Hence the pair of Eqs. (77) and (78) imply Eq. (75). Eq. (75) in turn
implies that the order $r$ of any $n$ modulo $N=pq$ is $\leq \phi (N)/2,$ so
that $r$ indeed is 
\mbox{$<$}%
$N/2,$ an inequality that is cruical to the derivation of the important Eq.
(57).

The probability that a randomly selected positive integer 
\mbox{$<$}%
$N$ will be coprime to $N$ obviously is 
\begin{equation}
\frac{\phi (N)}{N-1}=1-\frac{p-1+q-1}{N-1},
\end{equation}
using Eq. (74). For actual RSA key numbers, e.g., RSA-309, the right side of
Eq. (79) will be indistinguishable from unity for all practical purposes.
For instance if the smaller of $p$ and $q$ is not less than $N^{1/4},$ the
larger of $p$ and $q$ will be no greater than $N^{3/4},$ and the right side
of Eq. (79) is approximately 1 - $N^{1/4}$ which, even for a key number as
small as RSA-155, differs from unity by approximately (10)$^{-39}.$
Correspondingly for actual RSA key numbers the magnitude of $\phi (N)$ can
be taken equal to $N$ for all practical purposes. It is additionally worth
noting that because $\phi (N)$ evidently also equals the number of integers $%
i$ coprime to $N$ in the ranges $N+1<i<2N,$ 2$N+1<i<3N,$ etc., the
probability that a randomly selected integer 
\mbox{$<$}%
$N^{2}$ (or 
\mbox{$<$}%
2$^{y},$ with $y$ defined as under Subheading III.C.1) will be coprime to $N$
also can be taken equal to unity for all practical purposes.

\subsection{Proof the RSA System Correctly Deciphers.}

Eqs. (2) and (3), together with the definition of $N,$ imply 
\begin{equation}
u\equiv c^{ed}=(c^{e})^{d}\hspace{1in}(%
\mathop{\rm mod}%
\;pq).
\end{equation}
Because $d$ and e are positive integers by definition, and recalling the
definition of $\phi $ = $\phi (N),$ Eq. (4) implies 
\begin{equation}
de=1+z\phi =1+z(p-1)(q-1),
\end{equation}
where $z$ is a positive integer. Because by definition $u$ and $c$ are
positive integers 
\mbox{$<$}%
$N$ = $pq$, knowing $u\equiv c$ (mod $pq)$ implies $u=c.$ Thus to prove the
RSA system enables Bob to correctly decipher Alice's message I merely need
show that 
\begin{equation}
c^{1+z(p-1)(q-1)}\equiv c\hspace{1in}(%
\mathop{\rm mod}%
\;pq).
\end{equation}

If $c$ is coprime to $N$ then the now demonstrated Eq. (9) implies 
\begin{equation}
c^{z(p-1)(q-1)}\equiv 1\hspace{1in}(%
\mathop{\rm mod}%
\;pq),
\end{equation}
from which Eq. (82) immediately follows after multiplying both sides of Eq.
(83) by $c.$ If $c$ is not coprime to $pq,$ $c<N$ is divisible by one of $p$
and $q$ but not by both. Suppose for concreteness $c$ is divisible by $q,$
i.e., suppose $c=bq,$ $b$ a positive integer 
\mbox{$<$}%
$p.$ Then Eq. (76) holds for $a$ = $c$ and implies 
\begin{equation}
c^{z(p-1)(q-1)}\equiv 1\hspace{1in}(%
\mathop{\rm mod}%
\;p).
\end{equation}
But if $x-y$ is divisible by $p,$ then $q(x-y)$ is divisible by $qp,$
implying further that $bq(x-y)$ is divisible by $qp.$ Hence Eq. (84) implies 
\begin{equation}
bqc^{z(q-1)(p-1)}\equiv bq\hspace{1in}(%
\mathop{\rm mod}%
\;qp).
\end{equation}
Eq. (85) is Eq. (82) in the circumstance that $c=bq.$ Correspondingly Eq.
(82) will hold if $c$ is divisible by $p.$ We conclude that Eq. (82) holds
for every $c$ in Alice's cryptogram whether c is coprime to $N$ or not. This
completes the demonstration that the RSA system does enable Bob to correctly
decipher Alice's cryptogram.

\subsection{Classical Computer Calculations Pertinent to Shor Algorithm
Factorization.}

This Subsection discusses the various classical computer calculations
mentioned in earlier Subsections of this paper. The results under
Subheadings 1 through 3 below are the bases for the assertions made in
Subsection III.D about the growth with $N$ of the classical computer
calculations required to factor $N=pq$ using Shor's algorithm. How Bob can
determine his decryption exponent $d$ from $\phi $ = $\phi (N)$ and his
encryption exponent $e$ (recall Subsection II.A) is described under
Subheading 4 below.

\subsubsection{Greatest Common Divisors. The Euclidean Algorithm.}

A convenient method for computing the gcd of two positive integers was first
described by Euclid. My discussion of the Euclidean algorithm closely
follows Rosen.$^{68}$ Suppose the integer $x\geq 1$ is the gcd of the two
positive integers $s_{0}$ and $s_{1},$ where $s_{0}>s_{1}>1.$ The Euclidean
algorithm determines $x$ as follows. Divide $s_{0}$ by $s_{1},$ thereby
obtaining the remainder $s_{2}$ $\geq 0.$ By definition $s_{2}$ is 
\mbox{$<$}%
$s_{1}$ and satisfies 
\begin{equation}
s_{0}=z_{0}s_{1}+s_{2},
\end{equation}
where $z_{0}$ is a non-negative integer and 0 $\leq $ $s_{2}<s_{1}.$ Since $%
x $ is a divisor of $s_{0}$ and $s_{1},$ Eq. (86) implies $x$ is a divisor
of $s_{2}.$ Proceeding in this fashion, always dividing the previous divisor 
$s_{j}$ by the previous remainder $s_{j+1},$ one obtains a sequence of
remainders $s_{2},s_{3},...,s_{j+1},s_{j+2},...,$ each of which is a
multiple of $x.$ Moreover since each $s_{j}$ always is 
\mbox{$>$}%
$s_{j+1},$ the sequence eventually must terminate with some $s_{k+2}=0,$
i.e., eventually there will be the simple equation 
\begin{equation}
s_{k}=z_{k}s_{k+1}.
\end{equation}

It now can be seen that $s_{k+1}$ is not merely a multiple of $x,$ the gcd
of $s_{0}$ and $s_{1};$ rather $s_{k+1}=x.$ Eq. (87) shows $s_{k}$ is a
multiple of $s_{k+1}.$ The preceding equation in the series, namely 
\begin{equation}
s_{k-1}=z_{k-1}s_{k}+s_{k+1},
\end{equation}
then implies $s_{k-1}$ also is a multiple of $s_{k+1}.$ Thus, proceeding in
this fashion back through the series of equations which led from Eq. (86) to
Eq. (88), it can be concluded that both $s_{0}$ and $s_{1}$ are multiples of 
$s_{k+1}.$ Hence $s_{k+1}$ must be a divisor of $x,$ the gcd of $s_{0}$ and $%
s_{1}.$ But we already have shown that $x$ is a divisor of $s_{k+1}.$
Consequently $x$ must be identical with $s_{k+1},$ the last remainder before 
$s_{k+2}=0.$ If $s_{k+1}=1,$ then $s_{1}$ is coprime to $s_{0}.$

I will illustrate the use of the Euclidean algorithm to find the factor 5 of
N = 55 when, as explained at the end of Subsection III.A, for $n=12$ it is
deduced that $f_{2}+1=35$ must be divisible by one of the factors of 55. We
have: 55 = 1$\times 35+20;$ 35 = 1$\times 20+15;$ 20 = 1$\times 15+5;$ 15 = 3%
$\times 5+0.$ Therefore 5 is the gcd of 55 and 35. Similarly, suppose we had
decided (unnecessarily for large $N$ as is discussed under Subheading
III.C.3) to verify that 12 actually is coprime to 55. Now we have: 55 = 4$%
\times 12+7;$ 12 = 1$\times 7+5;$ 7 = 1$\times 5+2;$ 5 = 2$\times 2+1;$ 2 = 2%
$\times 1+0.$ So 1 is the gcd of 12 and 55, i.e., 12 really is coprime to 55.

How many classical computer bit operations are required to obtain the gcd of
two large numbers $N_{1}$ and $N_{2}$ via the Euclidean algorithm? Define $%
L_{1}=$ log$_{2}N_{1},$ $L_{2}=$ log$_{2}N_{2};$ as discussed immediately
beneath Eq. (29), for large $N_{1},N_{2}$ the quantities $L_{1},L_{2}$
differ negligibly from the number of digits in the binary expansions of $%
N_{1},N_{2}$ respectively. Then according to a theorem by Lam\'{e}$^{68}$
the number of divisions needed to find the gcd of $N_{1}$ and $N_{2}$ using
the Euclidean algorithm is at most $O(L_{1}),$ where $O$ is the $Order$ $of$
symbol used in Eq. (28). The number of bit operations in any one of those
divisions hardly can exceed the number of bit operations in the first of
those divisions (of $N_{1}$ by $N_{2}),$ wherein the dividend $N_{1}$ and
divisor $N_{2}$ are at their respective maximum values. Although at first
sight the number of bit operations required to divide $N_{1}$ by $N_{2}$ is$%
^{69}$ $O(L_{1}L_{2}),$ in actuality there exist$^{70}$ sophisticated
classical computer algorithms which for large $N_{1}$, $N_{2}$ reduce the
number of bit operations required for this division to $O[L_{1}($log$%
_{2}L_{1})($log$_{2}$log$_{2}L_{1})].$

Consequently the number of computer bit operations required to obtain the
gcd of two large numbers $N_{1}$ and $N_{2}$ using the Euclidean algorithm
surely is $O[L_{1}^{2}($log$_{2}L_{1})($log$_{2}$log$_{2}L_{1})].$ Returning
now to the discussion in Subsection III.D of the classical computer
calculations required for factoring using Shor's algorithm, the two numbers
whose gcd is required always will be no larger than $N$ = $pq$ and 1 + $%
f_{r/2},$ where according to Eq. (8) every $f_{j}$ is 
\mbox{$<$}%
$N$ by definition. Once an $r$ permitting factorization of $N$ has been
inferred [namely an $r$ satisfying Eqs. (11), (12) and (14)], only a single
gcd computation will be needed in order to complete the factorization of $N.$
No such gcd calculation is needed until a so usable $r$ has been inferred.
It follows that the number of bit operations required for the gcd
calculations involved in factoring $N$ = $pq$ using Shor's algorithm will
not grow faster with increasing $N$ than $O[L^{2}($log$_{2}L)($log$_{2}$log$%
_{2}L)],$ the same growth rate with $L$ as is given by Eq. (28).

\subsubsection{Continued Fraction Expansions.}

Rosen$^{60}$ explicitly demonstrates that the divisions performed in finding
the gcd of the positive integers $s_{0}$ and $s_{1}$ via the Euclidean
algorithm are the same as the divisions performed in constructing the
continued fraction expansion of the fraction $s_{1}/s_{0.}$ By way of
illustration, suppose we seek the gcd of the integers $2253$ and 4096 whose
ratio was expanded in the continued fraction of Eq. (55). We have: 4096 = 1$%
\times 2253+1843;$ 2253 = 1$\times 1843+410;$ 1843 = 4$\times 410+203;$ 410
= 2$\times 203+4;$ and so on. Evidently the divisions performed to obtain
these relations indeed are identical with those performed in constructing
the right side of Eq. (55). Thus to estimate the number of bit operations
required to compute the continued fraction convergents of any one $c/2^{y}$
measured as described under Subheading III.C.6, the result obtained under
the immediately preceding Subheading is immediately applicable. It is
necessary only to observe that for sufficiently large $N$ the value of $y$ =
log$_{2}2^{y}$ differs negligibly from 2$L=$ log$_{2}N^{2}.$ Accordingly,
the number of bit operations required to perform a typical continued
fraction expansion of a measured $c/2^{y}$ should be $O[(2L)^{2}($log$%
_{2}2L)($log$_{2}$log$_{2}2L)]=$ $O[L^{2}($log$_{2}L)($log$_{2}$log$_{2}L)],$
precisely the same result as obtained under the previous Subheading for the
Shor algorithm gcd calculation.

Unlike the gcd case, however, a continued fraction expansion is required
every time a $c/2^{y}$ is measured. The expected number of repetitions of
such measurements has been discussed under Subheading III.C.8 and in
Subsection III.D. Those discussions indicated that a probable overestimate
of the required number of repetitions is log$_{2}L,$ implying that the
overall number of bit operations required to perform the continued fraction
expansions during factoring by Shor's algorithm may grow with increasing $N$
as fast as, but no faster than, the right side of Eq. (59).

\subsubsection{Modular Exponentiation.}

Verifying that $n^{r}\equiv 1$ $(%
\mathop{\rm mod}%
\;N),$ and then computing $n^{r/2}\equiv f_{r/2}$ $(%
\mathop{\rm mod}%
\;N),$ so as hopefuly to factor $N=pq$ via Eq. (11), involves so-called {\it %
modular exp}$onentiation$. Volovich$^{54}$ sketches the proof that the
number of bit operations required to calculate $n^{j}$ $(%
\mathop{\rm mod}%
\;N)$ on a classical computer is $O[L^{2}($log$_{2}L)($log$_{2}$log$_{2}L)],$
i.e., grows with $N$ as does the right side of Eq. (28). As discussed under
Subheading III.C.8, a few repetitions of these exponentiations may be
necessary because the probability that a chosen $n$ will yield an $r$
permitting factorization of $N$ via Eq. (11) is only about 1/2. A few more
exponentiations may be necessary to rule out, as possible values of $r,$ the
denominators $b$ (and small multiples thereof) of convergents $a/b$ to
measured $c/2^{y}$ when $a/b=d/r$ but $b$ 
\mbox{$<$}%
\mbox{$<$}%
$r,$ also as discussed under Subheading III.C.8. It does not appear,
however, that as many repeated exponentiations ever will be required as the $%
O($log$_{2}L)$ repetition factor inferred from Eq. (58). Consequently, just
as under the immediately preceding Subheading, the number of bit operations
needed to perform the classical computer modular exponentiations that arise
during factorization via Shor's algorithm may grow with increasing $N$ as
fast as, but surely no faster than, the right side of Eq. (59).

\subsubsection{Finding the Decryption Exponent.}

We need to solve Eq. (4) for $d,$ knowing $e$ and $\phi .$ There is a known$%
^{71}$ closed formula for $\phi (\phi ),$ the totient function of $\phi $
(recall Subsection IV.B), in terms of the prime factors of $\phi .$ Thus if
we could factor $\phi $ we immediately could find $d.$ Namely since $e$ is
coprime to $\phi $ (recall Subsection II.A), Eq. (73) implies 
\begin{equation}
e^{\phi (\phi )}\equiv 1\hspace{1in}(%
\mathop{\rm mod}%
\;\phi ).
\end{equation}
Consequently Eq. (4) is solved by 
\begin{equation}
d\equiv e^{\phi (\phi )-1}\hspace{1in}(%
\mathop{\rm mod}%
\;\phi ).
\end{equation}

When $N$ is of the magnitude of modern RSA key numbers, factoring a $\phi
=(p-1)(q-1)\cong N$ (again recall Subsection IV.B) can be difficult, though
perhaps not as difficult as factoring $N=pq$ itself. In practice, therefore, 
$d$ probably would be determined as follows. Eq. (4) means there is an
integer $k$ such that 
\begin{equation}
ed=1+k\phi .
\end{equation}
Eq. (91) is a $Diophantine$ $equation$ in the unknowns $k$ and $d,$ whose
solution can be found$^{72}$ by working backwards from the set of equations
constituting the Euclidean algorithm for the gcd of $e$ and $\phi .$

I will illustrate this just explained method of solving Eq. (4) in our
oft-employed illustrative case $N=55$ We have $\phi =40,$ and have chosen $%
e= $ 23 (again recall Susection II.A). The Euclidean algorithm equations for
obtaining the gcd of 40 and 23 are: 40 = 1$\times 23$ + 17; 23 = 1$\times 17$
+ 6; 17 = 2$\times 6$ + 5; 6 = 1$\times 5$ + 1; 5 = 5$\times 1$ + 0,
verifying that our $e$ is coprime to our $N.$ Now, working backwards: 6 - 5
= 1; 5 = 17 - 2$\times 6,$ so 6 - (17 - 2$\times 6)$ = 3$\times 6$ - 17 = 1;
6 = 23 - 17, so 3$\times (23$ - 17) - 17 = 3$\times 23$ - 4$\times 17$ = 1;
17 = 40 - 23, so 3$\times 23$ - 4$\times (40$ - 23) = 7$\times 23-4\times 40$
= 1. This last equation is of the form of Eq. (91), and implies 7$\times 23$ 
$\equiv 1$ $(%
\mathop{\rm mod}%
\;40).$ Therefore the desired $d$ equals 7, as asserted at the close of
Subsection II.A.

It is apparent that the computing effort required of Bob in determining his $%
d$ via this just described procedure is utterly negligible compared to the
computing effort he will endure in decrypting the many messages he expects
to receive from Alice.

.

-----------------------------------------------------------------------------

$^{1}$ {\footnotesize N. D. Mermin, ''From Cbits to Qbits: Teaching computer
scientists quantum mechanics,'' Am. J. Phys. {\bf 71}, 23 (2003).}

$^{2}$ {\footnotesize L. K. Grover, ''From Schrodinger's Equation to the
quantum search algorithm,'' Am. J. Phys. {\bf 69}, 769 (2001)}{\small .}

$^{3}$ {\footnotesize P. W. Shor, ''Algorithms for quantum computation:
discrete logarithms and factoring,'' in {\it Proceedings} {\it of the 35
Annual Symposium on the Foundations of Computer Science, }edited by S.
Goldwasser (IEEE Computer Society Press, Los Alamitos, CA), pp. 124-134
(1994). P.W. Shor, SIAM J. Comp. {\bf 26}, 1474 (1997), first published as
quant-ph/9508027 (1/25/96), provides an expanded version of Shor's original
paper.}

$^{4}$ {\footnotesize Cf.,} {\footnotesize e.g., I. V. Volovich, ''Quantum
Computing and Shor's Factoring Algorithm,'' quant-ph/0109004 (9/2/01); A.
Ekert and R. .Josza, ''Quantum computation and Shor's factoring algorithm,''
Rev. Mod. Phys. {\bf 68}, 733 (1996)}{\bf ; }{\footnotesize M. A. Nielsen
and I. L. Chuang, {\it Quantum Computation and Quantum Information }
(Cambridge 2000), pp. 232-247.}

$^{5}$ {\footnotesize Cf., e.g., C. P. Williams and S. H. Clearwater, {\it %
Explorations in Quantum Computing }(Springer 1998), pp. 130-145; G. Johnson, 
{\it A Shortcut Through Time}} {\footnotesize (Knopf 2003), pp. 66-82; J.
Brown, {\it The Quest for the Quantum Computer} (Simon \& Schuster 2000),
pp. 170-188.}

$^{6}$ {\footnotesize Cf., e.g., arXiv.org/archive/quant-ph. See also
www.eg.bucknell.edu/\symbol{126}dcollins/research/qcliterature.html,
maintained by David Collins at his present place of employment.}

$^{7}${\footnotesize \ An exposition (suitable for the non-specialist
readers of this journal) of what is now termed the RSA public key system can
be found in Williams and Clearwater, {\it ibid,} pp. 122-127. An important
reference, probably useful to cryptography specialists only however, is A.
Menezes, P. van Oorschot and S. Vanstone, {\it Handbook of Applied
Cryptography }(CRC Press 1996), Chapter 8. The RSA system was first proposed
by R. Rivest, A. Shamir and L. Adleman, ''On Digital Signatures and
Public-Key Cryptosystems,'' MIT Laboratory for Computer Science Technical
Report MIT/LCS/TR-212 (January 1979).}

$^{8}$ {\footnotesize A. Ekert, ''From quantum code-making to quantum
code-breaking,'' quant-ph/9703035 (3/19/97).}

$^{9}$ {\footnotesize D. Kahn,{\it \ The Codebreakers: The Story of Secret
Writing }(Scribner 1996); for the definitions adopted herein see pp.
xv-xviii and 989.}

$^{10}\ ${\footnotesize S. Singh, {\it The Code Book }(Doubleday 1999), esp.
chs. 6 and 7.}

$^{11}\ ${\footnotesize N. Gisin {\it et al, }''Quantum cryptography,'' Rev.
Mod. Phys. {\bf 74}, 145 (2002), at pp. 147-8; Singh, {\it ibid, }pp.
268-273. }

$^{12}${\footnotesize {\it \ }Kahn, {\it ibid, }pp. 71-88.}

$^{13}$ {\footnotesize Kahn, {\it ibid, }pp. 93-105.}

$^{14}\ ${\footnotesize ''The} {\footnotesize Gold-Bug,'' published in 1843.
See}$,${\footnotesize \ e.g., {\it Complete Stories and Poems of Edgar Allan
Poe }(Doubleday 1966), pp. 70 and 819. }

$^{15}$ {\footnotesize ''The Adventure of the Dancing Men.'' See}$,$%
{\footnotesize \ e.g., {\it The Complete Sherlock Holmes}} {\footnotesize %
(Doubleday 1966), p. 593. {\it \ }}

$^{16}$ {\footnotesize For instance, the {\it Pittsburgh Post Gazette.}}

$^{17}$ {\footnotesize Singh, {\it ibid, }pp. 20-25, and Kahn, {\it ibid, }
pp. 99-105, provide detailed illustrative cryptanalyses of such cryptograms. 
}

$^{18}$ {\footnotesize Singh, {\it ibid, }ch. 4, describes the Enigma
machine and recounts the remarkable story of how its cryptograms were
cryptanalysed. See also A. Hodges, {\it Alan Turing: The Enigma }(Simon and
Schuster 1983), ch. 4. }

$^{19}$ {\footnotesize Cf., e.g., Ekert, {\it ibid.} }

$^{20}$ {\footnotesize Actually it is possible, though intrinsically
inconvenient, for Alice and Bob to establish a secure key via conventional
communication channels without meeting, as was discovered in 1976; see
Singh, {\it ibid, }pp. 253-267. Secure key distribution also is possible (in
theory at least) via ''quantum channels'', e.g., channels that carry pairs
of spin 1/2 particles whose spin orientations can be measured by Alice and
Bob; see Ekert, {\it ibid}. These secure key distribution schemes are beyond
the scope of this paper.}

$^{21}$ {\footnotesize I do not pretend that this analogy between
cryptographic keys and safes is original. See, e.g., Gisin, {\it ibid}. }

$^{22}$ {\footnotesize Singh, {\it ibid, }pp. 245-249 and 379.}

$^{23}$ {\footnotesize See the website maintained by Jim Price at
www.jimprice.com/jim-asc.htm, esp. the link to} {\footnotesize a
decimal-to-ASCII chart.}

$^{24}$ {\footnotesize Cf. any textbook on elementary number theory, e.g.,
K. H. Rosen, {\it Elementary Number Theory and Its Applications }
(Addison-Wesley 1993), pp. 119-125.}

$^{25}$ {\footnotesize Menezes, van Oorschot and Vanstone, ibid (see fn. 7
supra), esp. p. 292.}

$^{26}$ {\footnotesize ''How large a key should be used in the RSA
cryptosystem?'' (RSA Security 2004), at www.
rsasecurity.com/rsalabs/faq/3-1-5.html; ''TWIRL and RSA Key Size'' (RSA
Security 2003), at www. rsasecurity.com/technotes/twirl.html..}

$^{27}$ {\footnotesize A. Ekert, ''Quantum Cryptoanalysis-Introduction'' (as
updated by Wim van Dam, Centre for Quantum Computation, June 1999), at
www.qubit.org/intros/cryptana.html.}

$^{28}$ {\footnotesize R. Roskies, Scientific Director Pittsburgh
Supercomputing Center, private communication.}

$^{29}$ {\footnotesize Based on measurements by the WMAP satellite. Cf.
http://map.gsfc.nasa.gov/m\_uni/uni\_101age.html.}

$^{30}$ {\footnotesize Brown, {\it ibid }(see fn. 5 {\it supra), }pp.
164-166.}

$^{31}$ {\footnotesize R. Crandall and C. Pomerance, {\it Prime Numbers. A
Computational Perspective }(Springer 2001){\it , }pp. 225-232.}

$^{32}$ {\footnotesize ''What is the RSA Factoring Challenge?'' (RSA
Security 2004), at www. rsasecurity.com/rsalabs/faq/2-3-6.html.}

$^{33}$ {\footnotesize Crandall and Pomerance, {\it ibid, }pp. 242-258.}

$^{34}$ {\footnotesize ''What are the best factoring methods in use today?''
(RSA Security 2004), at www.rsasecurity.com/rsalabs/faq/2-3-4.html.}

$^{35}$ {\footnotesize ''Factorization of RSA-160'' (RSA Security 2004), at
www.rsasecurity.com/rsalabs/faq/challenges/factoring/rsa160.html (to verify
the April 1, 2003 date, click on the link to ''Paul Zimmerman's factoring
page'').}

$^{36}$ {\footnotesize \ ''The RSA Challenge Numbers'' (RSA Security 2004),
at www.rsasecurity.com/rsalabs/challenges/factoring/numbers.html.}

$^{37}$ {\footnotesize ''TWIRL and RSA Key Size'' {\it ibid } (see fn. 26 
{\it supra} ).}

$^{38}$ {\footnotesize Crandall and Pomerance, {\it ibid, }p. 265. }

$^{39}$ {\footnotesize N. Koblitz, {\it A Course in Number Theory and
Cryptography\ (}Springer 1987), pp. 3-4.}

$^{40}$ {\footnotesize Williams and Clearwater, {\it ibid } (see fn. 5 {\it %
supra} ), p. 35.}

$^{41}$ {\footnotesize P. Ribenboim, {\it The New Book of Prime Number
Records (}Springer 1995), p. 156, writes: ''It is fairly easy, in practice,
to produce large primes. It is, however, very difficult to produce a
theoretical justification for the success of the method.'' See also the same
author's ''Selling Primes,'' Mathematics Magazine {\bf 68, }175 (1995). The
essential point is that finding the primes {\it p} and {\it q} which will be
multiplied to construct {\it N} can be accomplished in computing times at
most polynomial in {\it L }= log}$_{_{2}}${\footnotesize {\it N,} whereas
factoring {\it N }to find its prime factors {\it p} and {\it q }requires
computing times subexponential in {\it L (}as we have discussed, assuming
only classical comuters are available).. }

$^{42}$ {\footnotesize A recent test run demonstrated that even with an RSA
key number of 2048 binary bits (i.e., an RSA-617) a message consisting of
approximately 32,000 ASCII characters could be routinely enciphered and
deciphered in times of the order of seconds and at most minutes
respectively, employing merely a 700 Megahertz desktop computer (hardly a
supercomputer). For example, using block sizes of 52 ASCII characters
(recall Subsection II.C), the encryption and decryption times were 1.46 and
30.3 seconds respectively. Sam Scheinman, software engineer consultant,
private communication. }

$^{43}$ {\footnotesize Rosen, {\it ibid }(see fn. 24 {\it supra), }pp. 278-9.%
}

$^{44}$ {\footnotesize Crandall and Pomerance, {\it ibid, }p. 386.}

$^{45}$ {\footnotesize Koblitz, {\it ibid, }p. 94.}

$^{46}${\footnotesize \ A. Odlyzko, ''Discrete Logarithms: The Past and the
Future,'' Designs, Codes and Cryptography {\bf 19, }59 (2000).}

$^{47}$ {\footnotesize J. Eisen and M. Wolf, ''Quantum Computing,''
quant-ph/0401019 (1/5/04). Cf, esp., p. 16.}

$^{48}$ {\footnotesize Cf., e.g., D. J. Griffiths,{\it \ Introduction to
Quantum Mechanics (}Prentice Hall 1994), or S. Gasiorowicz, {\it Quantum
Physics, }3d edn (Wiley 2003).}

$^{49}\ ${\footnotesize \ Cf., e.g., Nielsen and Chuang, {\it ibid } (see
fn. 4 {\it supra), }ch. 7. }

$^{50}$ {\footnotesize Cf., e.g., Griffiths, {\it ibid, }pp. 154-159, or V.
Scarani, ''Quantum Computing,'' Am. J. Phys. {\bf 66}, 956 (1998). }

$^{51}$ {\footnotesize Griffiths, {\it ibid, }p. 12 and Gasiorowicz, {\it %
ibid, }p. 139. }

$^{52}$ {\footnotesize Mermin, {\it ibid } (see fn. 1 {\it supra), }esp. his
Eq. (35).}

$^{53}$ {\footnotesize Cf., e.g., Nielsen and Chuang, {\it ibid, }ch. 4.}

$^{54}$ {\footnotesize Volovich, {\it ibid } (see fn. 4 {\it supra)}.}

$^{55}$ {\footnotesize Crandall and Pomerance, {\it ibid, }p. 7.}

$^{56}$ {\footnotesize Williams and Clearwater, {\it ibid } (see fn. 5 {\it %
supra), }esp. pp. 136-7.}

$^{57}$ {\footnotesize Cf., e.g., Nielsen and Chuang, {\it ibid, }pp. 18-19.}

$^{58}${\footnotesize \ Nielsen and Chuang, {\it ibid, }pp. 194-198.}

$^{59}$ {\footnotesize Nielsen and Chuang, {\it ibid, }pp. 217-220.}

$^{60}$ {\footnotesize Rosen, {\it ibid, }pp. 394-403.}

$^{61}$ {\footnotesize Cf., e.g., A. Ekert and R. .Josza, {\it ibid } (see
fn. 4 {\it supra). }These authors refer to G. H. Hardy and E. M. Wright, 
{\it An Introduction to the} {\it Theory of Numbers (}Clarendon, Oxford
1965), Sec. 10.15, for a proof of the theorem.}

$^{62}$ {\footnotesize Crandall and Pomerance, {\it ibid, }p. 11.}

$^{63}$ {\footnotesize Ribenboim, {\it ibid }(see fn. 4 {\it supra), }pp.
319-320.}

$^{64}$ {\footnotesize D. E. Knuth, {\it The Art of Computer Programming, }%
3d edn{\it \ (}Addison-Wesley 1981), vol. 2, pp. 290 and 300 (see problem 8).%
}

$^{65}$ {\footnotesize L. Vandersypen {\it et al.} , ''Experimental
realization of Shor's quantum factoring algorithm using nuclear magnetic
resonance,'' Nature {\bf 414, }883 (2001).{\bf \ }}

$^{66}$ {\footnotesize Rosen, {\it ibid, }pp. 201-204. }

$^{67}$ {\footnotesize Rosen, {\it ibid, }p. 187; L. E. Dickson, {\it Modern
Elementary Theory of Numbers }(Univ. of Chicago 1939), p. 12, quotes the
date of Fermat's Little Theorem.}

$^{68}$ {\footnotesize Rosen, {\it ibid, }pp. 80-84.}

$^{69}$ {\footnotesize Koblitz, {\it ibid, }pp. 6-7.}

$^{70}$ {\footnotesize Rosen, {\it ibid, }p. 61; Knuth, {\it ibid, }p. 295.}

$^{71}$ {\footnotesize Rosen, {\it ibid, }p. 210.}

$^{72}$ {\footnotesize Rosen, {\it ibid, }pp. 132-3.}

\end{document}